\documentclass[12pt]{iopart}
\usepackage{iopams}  
\usepackage{setstack}

\begin{document}
\title[Integrability conditions for two-dimensional lattices]{Integrability conditions for two-dimensional lattices}

\author{I T Habibullin$^{1,2}$, M N Kuznetsova$^1$ and A U Sakieva$^3$}

\address{$^1$Institute of Mathematics, Ufa Federal Research Centre, Russian Academy of Sciences,
112 Chernyshevsky Street, Ufa 450008, Russian Federation}
\address{$^2$Bashkir State University, 32 Validy Street, Ufa 450076 , Russian Federation}
\address{$^3$State Budgetary Educational Institution,
	G. Almukhametov Republican Boarding School, 25/1 Zorge Street, Ufa 450059, Russian Federation}
\eads{\mailto{habibullinismagil@gmail.com}, \mailto{mariya.n.kuznetsova@gmail.com} and \mailto{alfya.sakieva@yandex.ru}}

\begin{abstract}
In the article some algebraic properties of nonlinear two-dimensional lattices of the form $u_{n,xy} = f(u_{n+1}, u_n, u_{n-1})$ are studied. The problem of exhaustive description of the integrable cases of this kind lattices remains open. By using the approach, developed and tested in our previous works we adopted the method of characteristic Lie-Rinehart algebras to this case. In the article we derived an effective integrability conditions for the lattice and proved that in the integrable case the function $f(u_{n+1}, u_n, u_{n-1})$ is a quasi-polynomial satisfying the following equation
$\frac{\partial^2}{\partial u_{n+1}\partial u_{n-1}}f(u_{n+1}, u_n, u_{n-1})=Ce^{{\alpha}u_n-{\frac{{\alpha}m}{2}}u_{n+1}-{\frac{{\alpha}k}{2}}u_{n-1}},$
where $C$ and $\alpha$ are constant parameters and $k,\,m$ are nonnegative integers.  
\end{abstract}

\maketitle

\eqnobysec

\section{Introduction}

Multidimensional integrable equations, such as the KP and Davey-Stewartson equations, the two-dimensional Toda lattice and so on have important applications in physics and geometry. In recent years, various methods have been developed to study such kind equations (see, for instance, \cite{Bogdanov}-\cite{Zakharov}).

In this paper we consider a class of the  lattices of the form
\begin{equation}  \label{eq_main}  
u_{n,xy} = f(u_{n+1}, u_n, u_{n-1}), \quad -\infty < n < \infty,
\end{equation}
where the unknown $u_n = u_n(x,y)$ depends on the real $x,y$ and the integer $n$. The function  $f$ of three variables is assumed to be analytic in a domain  $D\subset \mathbb{C}^3$. 

Recall that class (\ref{eq_main}) contains such a famous equation as the two-dimensional Toda lattice, which can be represented as (\ref{eq_main}) in the following three ways
\begin{eqnarray}  \label{Toda1}
u_{n,xy}& =& e^{u_{n+1} - 2 u_n + u_{n-1}},\\
v_{n,xy} &=& e^{v_{n+1}-v_n} - e^{v_n - v_{n-1}},  \label{Toda2}\\
  \label{Toda3}
w_{n,xy}& =& e^{w_{n+1}} - 2 e^{w_{n}} + e^{w_{n-1}}.
\end{eqnarray}
The lattices are related to each other by the linear substitutions $w_n=v_{n+1}-v_n$, $v_n=u_{n}-u_{n-1}$. 

An important open problem is to describe all of the integrable cases in the class (\ref{eq_main}). The purpose of the present article is to give an appropriate algebraic formalization of the classification problem and derive effective necessary integrability conditions for (\ref{eq_main}). Our investigation is based on the classification scheme, outlined in \cite{H2013}-\cite{HabKuznetsova19}. The scheme realizes the commonly accepted opinion, that any integrable equation in 3D admits a large set of integrable in a sense, two dimensional reductions. The hydrodynamic reduction method, developed in \cite{Tsarev}-\cite{Ferapontov2006} is one of the fruitful applications of this idea. According to the method 3D equation is integrable if it admits sufficiently large class of 2D reductions in the form of the integrable hydrodynamic type systems. In our study we observed that existence of a sequence of Darboux integrable reductions might be regarded as a sign of integrability as well.

In the case of lattice (\ref{eq_main}) we use the following 

{\bf Definition 1.} {\it Lattice (\ref{eq_main}) is called integrable if there exist locally analytic functions $\varphi$ and $\psi$ of two variables  such that for any choice of the integers $N_1$, $N_2$ the hyperbolic type system 
\begin{eqnarray}
&&u_{N_1,xy} = \varphi(u_{N_1+1},u_{N_1}), \nonumber \\
&&u_{n,xy}=f(u_{n+1}, u_{n},u_{n-1}),\qquad N_1 < n < N_2, \label{finite_sys} \\
&&u_{N_2,xy}=\psi(u_{N_2},u_{N_2-1}). \nonumber 
\end{eqnarray} 
obtained from lattice (\ref{eq_main}) is integrable in the sense of Darboux.}

The above integrable models (\ref {Toda1})-(\ref {Toda3}) are certainly integrable in the sense of our Definition 1 as well.
Recall that Darboux integrability  means that the system admits a complete set of nontrivial integrals in both characteristic directions of $x$ and $y$. An effective criterion of such kind of integrability is formulated in terms of the characteristic Lie-Rinehart algebras (see Theorem 1 below in \S 3). 
Properties of the characteristic algebra and its application for 1+1 dimensional continuous and discrete Darboux integrable systems are studied in \cite{ZMHS-UMJ}-\cite{Sakieva}. In the context of 2+1 dimensional models these algebras for the first time have been considered in \cite{Sh1995}.  In our recent works \cite{H2013}-\cite{HabKuznetsova19} we studied the classification problem for two-dimensional lattices of a special kind, where the method of Darboux integrable reductions and the algebraic approach were used as basic implements.  

We note that discussion on the alternative approaches to the study of Darboux integrable equations based on the higher symmetries, the Laplace invariants,  etc. can be found in  \cite{Anderson}-\cite{Smirnov}.

The article is organized as follows. In \S 2 we recall necessary definitions, introduce the notion of the characteristic Lie-Rinehart algebra and formulate an algebraic criterion of the Darboux integrability. In the third section we study the connection between the structure of the characteristic algebra and the properties of the function $f(u_{n+1},u_{n},u_{n-1})$ assuming that the reduced system (\ref{finite_sys}) is integrable in the sense of Darboux. We prove that $f$ is a quasi-polynomial with respect to all three arguments and give the complete description of its second order derivative $f_{u_{n+1},u_{n-1}}$. In the last fourth section we derive the main result of the article by formulating the necessary condition of integrability for the lattice (\ref{eq_main}). Some tediously long proofs carried over to the Appendix.

\section{Preliminaries}

Now let us recall some basic notions of the integrability theory. A function $I=I(x,\bar u,\bar u_x,\bar u_{xx},...)$ depending on a finite number of the dynamical variables is called an $y$-integral of system (\ref{finite_sys}) if it solves equation $D_yI=0$. Here $\bar u$ is a vector $\bar u=(u_{N_1}, u_{N_1 + 1}, \dots, u_{N_2})$, $\bar u_x$ is its derivative and so on. Similarly, a function $J=J(y,\bar u,\bar u_y,\bar u_{yy},...)$ is a $x$-integral if the equation  $D_xJ=0$ holds.
Integrals of the form $I=I(x)$ and $J=J(y)$, depending only on $x$ and $y$ are called trivial. A system (\ref{finite_sys}) is called  integrable in the sense of Darboux if it admits a complete set of functionally independent nontrivial integrals in both characteristic directions $x$ and $y$. Actually it is required that the number of functionally independent integrals is $N_1 + N_2 - 1$ in each direction.

Let us take a nontrivial $y$-integral $I=I(\bar u,\bar u_x,\bar u_{xx},...)$ of the system (\ref{finite_sys}). Obviously the operator $D_y$ acts on $I$ as a vector field of the following form: 
\begin{equation}\label{oper}
D_y=\sum_{j=N_1}^{N_2}\left(u_{j,y}\frac{\partial}{\partial{u_j}}+f_{j}\frac{\partial}{\partial{u_{j,x}}}+D_x(f_j)\frac{\partial}{\partial{u_{j,xx}}}+\cdots \right),
\end{equation}
where $f_j=f(u_{j+1},u_j,u_{j-1})$. The vector field (\ref{oper}) in a natural way splits down into a linear combination of the independent operators $X_j$ and $Z$
\[
D_y = \sum_{j=N_1}^{N_2}  u_{j,y} X_j  +Z,
\]
where
\begin{equation} \label{eq6}
X_j = \frac{\partial}{\partial u_j}, \quad Z = \sum_{j=N_1}^{N_2} \left( f_j \frac{\partial}{\partial u_{j,x}} + D_x(f_j) \frac{\partial}{\partial u_{j,xx}}+\cdots\right).
\end{equation}
Obviously equation $D_yI=0$ right away implies that $X_jI=0$ and $ZI=0$.

We study Darboux integrable systems by using characteristic algebras. Denote by $A$ the ring of locally analytic functions of the dynamical variables  $\bar u,\bar u_x,\bar u_{xx},\dots$. Let us introduce the Lie algebra $L_y$ with the usual operation $[W_1, W_2]=W_1 W_2-W_2 W_1$, generated by the differential operators $Z$ and $X_j$ defined in (\ref{eq6}) over the  ring $ A $. We assume the consistency conditions:
\begin{itemize}
\item[1).] $[W_1,aW_2]=W_1(a)W_2+a[W_1,W_2]$,
\item[2).] $(aW_1)b=aW_1(b)$
\end{itemize} 
to be valid for any $W_1,W_2\in L_y$ and $a,b\in A$. In other words we request that, if $W_1\in L_y$ and $a\in A$ then $aW_1\in L_y$. In such a case the algebra $L_y$ is called the Lie-Rinehart algebra \cite{Rinehart}, \cite{Million}. We call it also the characteristic algebra in the direction of $y$. In a similar way the characteristic algebra $L_x$ in the direction of $x$ is defined.

The algebra $L_y$ is of a  finite dimension if it admits a  basis consisting of a finite number of the elements $Z_1,Z_2,\dots,Z_k\in L_y$ such that an arbitrary operator $Z\in L_y$ is represented as a linear combination of the form
\[
Z=a_1Z_1+a_2Z_2+\dots a_kZ_k,
\]
where the coefficients are functions $a_1,a_2,\dots,a_k\in A$.

Now we are ready to formulate an algebraic criterion of the integrability of a hyperbolic type system in the sense of Darboux \cite{ZMHSbook, ZhiberK}, which plays a crucial role in our investigation.

{\bf Theorem 1.} \textit{System (\ref{finite_sys}) admits a complete set of the $y$-integrals (a complete set of the $x$-integrals) if and only if its characteristic algebra $L_y$ (respectively, its characteristic algebra $L_x$) is of a finite dimension.}

{\bf Corollary of Theorem 1.} {\it System (\ref{finite_sys}) is integrable in the sense of Darboux if and only if both characteristic algebras $L_x$ and $L_y$ are of a finite dimension.}

\section{Investigation of the characteristic algebras}

Assume that lattice (\ref{eq_main}) is integrable in the sense of Definition 1. Then for any pair of the integers $N_1,N_2$ hyperbolic type system (\ref{finite_sys}) has to be Darboux integrable. Therefore according to Theorem 1 the algebras $L_x$ and $L_y$ must be of a finite dimension. Since the elements of the algebras are vector fields with an infinite number of components, the problem of clarifying the linear dependence of a set of elements becomes very non-trivial. To this aim the following lemma provides a very useful implement \cite{ZMHS-UMJ}, \cite{ZMHSbook}.

{\bf Lemma 1.} \textit{If the vector field of the form
\[
Z = \sum_{i=N_1}^{N_2} z_{1,i} \frac{\partial}{\partial u_{i,x}} + z_{2,i} \frac{\partial}{\partial	u_{i,xx}} + \cdots
\]
solves the equation $\left[ D_x, Z \right] = 0$, then $Z=0$.}

 
{\bf Lemma 2.} \textit{The following formulas hold:}
\[
[D_x,X_j]=0,\quad  [D_x,Z]=-\sum _{j=N_1}^{N_2} f_j X_j.
\]

{\bf Proof of Lemma 2.} The operator $D_y$ acts on the function  $F(\bar u,\bar u_x,\bar u_{xx},\dots)$ by the following rule
\begin{equation*}
D_y=Z+\sum_{j=N_1}^{N_2}u_{j,y}X_j.
\end{equation*}
Since the operators $D_x$, $D_y$ commute with one another, we have the relation:
\begin{equation*}
[D_x, D_y]=[D_x,Z+\sum_{j=N_1}^{N_2}u_{j,y}X_j]=0.
\end{equation*}
Using properties of the commutators, we get 
$$[D_x,Z] + \sum_{j=N_1}^{N_2}f_j X_j + \sum_{j=N_1}^{N_2} u_{j,y} [D_x, X_j] =0.$$
By comparing the coefficients in front of the independent variables $u_{j,y}$, we easily obtain the statement of the Lemma 2.

Let us construct a sequence of the operators $Z_0$, $Z_1$, $Z_2, \ldots$ by setting
\begin{equation*}
Z_0 = \left[ X_0, Z \right], \quad Z_1 = \left[ X_0, Z_0 \right], \ldots, Z_{j+1} = \left[ X_0, Z_j \right], \ldots
\end{equation*}
It is easy to check that the relations hold
\begin{equation}  \label{DxZj}
\left[ D_x, Z_j \right] = -\sum^1_{k = -1} X^{j+1}_0 (f_k) X_k,
\end{equation}
where $f_0 = f(u_1, u_0, u_{-1})$.

{\bf Lemma 3.} \textit{Suppose that lattice (\ref{eq_main}) is integrable in the sense of Definiton 1, then the function $f = f(u_{1}, u_0, u_{-1})$ is a quasi-polynomial with respect to any of its arguments $u_{-1}$, $u_0$, $u_1$.}

{\bf Proof.} We suppose that lattice (\ref{eq_main}) is integrable in the sense of Definition 1, then the characteristic algebra should be finite-dimensional. That is why there exists a natural  $M$ such that $Z_{M+1}$ is linearly expressed through the previous members of the sequence:
\begin{equation}\label{mmm}
 Z_{M+1}=\sum^{M}_{i=0}{{\lambda}_i{Z_{i}}},
 \end{equation}
 where the operators  $Z_1, Z_2, \ldots Z_M$ are linearly independent.  
We commute both sides of equality (\ref{mmm}) with the operator $D_x$ and due to (\ref{DxZj}) we arrive at:
\begin{equation*}
 -\sum^{1}_{j=-1}X^{M+2}_0(f_j)X_j=\sum^{M}_{i = 0}\left\{D_x({\lambda}_i)Z_{i}-{\lambda}_i\sum^{1}_{k=-1}X^{i+1}_{0}(f_k)X_k\right\}.
 \end{equation*}
Comparing the coefficients before linearly independent operators $Z_i$ for $i=0,1,...,M$ one gets $D_x({\lambda}_i)=0$ and therefore ${\lambda}_i=const$. Comparison of the factors before $X_j$ yields for $j=-1,0,1$:
 \begin{equation*}\label{pro}
 (X^{M+1}_{0}-{\lambda}_M{X^M}-{\lambda}_{M-1}X^{M-1}-...-{\lambda}_0)X_0(f_j)=0.
 \end{equation*}
 Thus all three functions $f_{-1}=f(u_0,u_{-1},u_{-2})$, $f_0=f(u_{1},u_0,u_{-1})$ and $f_1=f(u_2,u_{1},u_0)$ are quasi-polynomials on the variable $u_0$, hence evidently $f(u_{1},u_0,u_{-1})$ is a quasi-polynomial with respect to all of its arguments. Lemma 3 is proved.

{\bf Lemma 4.} \textit{Operator  $W_0=\left[X_{1}, \left[ X_{-1}, Z \right]\right]$ satisfies the condition}
 \begin{equation*}	\label{eq9}
  \left[D_x, W_0\right]=-f_{u_1u_{-1}}X_0.
 \end{equation*}

{\bf Proof.}  Lemma  4 is easily proved by using the Jacobi identity and formula
\begin{equation*}
\left[D_x, \left[X_1, Z \right]  \right] = -\sum^2_{k =0} X_1(f_j)X_j.
\end{equation*}

Let us now construct a sequence of the form:
\begin{equation}		\label{seq_W}
W_0, \quad W_1 = \left[X_0, W_0  \right], \quad W_2 = \left[X_0, W_1  \right], \ldots, W_{k+1} = \left[X_0, W_k\right],...
\end{equation}
Elements of the sequence satisfy the formulas: 
\begin{equation} \label{DxWk}
\left[ D_x, W_k  \right] = -X^{k}_0(f_{u_1 u_{-1}})X_0.
\end{equation}
Since the characteristic algebra is finite-dimensional there exists a natural $M$ such that $W_{M+1}$ is linearly expressed through the previous members:
\begin{equation*}
W_{M+1} +  \lambda_M W_M + \cdots + \lambda_1 W_1 + \lambda_0 W_0 = 0,
\end{equation*}
where $W_0, \ldots, W_M$ are linearly independent. We commute both sides of this equality with the operator $D_x$ and apply formula (\ref{DxWk}). Thus we obtain the relations $D_x(\lambda_j)=0$ satisfied  for $j=0,1,...,M$ and an equation
\begin{equation*}		
X^{M+1}_0(f_{u_1 u_{-1}}) X_0 + \lambda_M X^M_0(f_{u_1 u_{-1}})X_0 + \cdots  + \lambda_0 f_{u_1 u_{-1}} X_0 = 0.
\end{equation*}
Obviously the latter implies:
\begin{equation*}		\label{char_pol1}
\left(X^{M+1}_0  + \lambda_M X^M_0 + \cdots  + \lambda_0 \right) f_{u_1 u_{-1}} = 0.
\end{equation*}
Let us denote through $\Lambda(\lambda)$ the characteristic polynomial of this linear ordinary differential equation, i.e. 
\begin{equation}\label{Lambda_00}
\Lambda(\lambda):=\lambda^{M+1}  + \lambda_M \lambda^M + \cdots  + \lambda_0. 
\end{equation}
Then we have that the differential operator $\Lambda(X_0)$ turns the function $g = f_{u_1 u_{-1}}$ to zero:
\begin{equation}		\label{Lambda_0}
\Lambda(X_0)g(u_{1}, u_0, u_{-1}) = 0
\end{equation}
and there is no any operator of lower order which annulates $g$.

Further it will be convenient to use the following notation for the commutator of two operators: $ad_X Y = \left[X, Y \right]$. In terms of this new notation, members of the sequence (\ref{seq_W}) are written as:
\begin{equation*}
W_0, \quad ad_{X_0} W_0, \quad ad^2_{X_0} W_0, \quad \ldots, \quad ad^{k+1}_{X_0} W_0.
\end{equation*}
Formula (\ref{DxWk}) takes the form:
\begin{equation}	\label{DxWk_new}
\left[ D_x, ad^k_{X_0}W_0 \right] = -X^{k}_0(f_{u_1 u_{-1}})X_0.
\end{equation}

{\bf Lemma 5.} \textit{Assume that the characteristic polynomial $\Lambda(\lambda)$ admits two different roots $\lambda = \alpha$ and $\lambda = \beta$. Then either a) $\alpha = - \beta$ or b) $\alpha = - 2 \beta$.}

{\bf Proof.} Let us construct polynomials  $\Lambda_{\alpha} (\lambda)$ and $\Lambda_{\beta}(\lambda)$ by the following rule:
\begin{equation*}
\Lambda_{\alpha} (\lambda)= \frac{\Lambda(\lambda)}{\lambda - \alpha}, \quad \Lambda_{\beta}(\lambda) = \frac{\Lambda(\lambda)}{\lambda - \beta}.
\end{equation*}
Then the operators 
\begin{equation*}
P_{\alpha}=\Lambda_{\alpha}(ad_{X_0}W_0), \quad P_{\beta}=\Lambda_{\beta}(ad_{X_0}W_0)
\end{equation*}
satisfy the relations
\begin{equation}	\label{lop}
 \left[D_x,P_{\alpha}\right]=A(u_{1},u_{-1})e^{\alpha{u_0}}X_0, \qquad \left[D_x,P_{\beta}\right]=B(u_{1},u_{-1})e^{\beta{u_0}}X_0,		
\end{equation}
where functions $A$ and $B$ don't vanish identically. These formulas are easily proved, let us begin the first one. The operation of commutation with $D_x$ by virtue of (\ref{DxWk_new}) satisfies the formula:
\begin{equation}\label{ss}
 \left[D_x,\Lambda_{\alpha}(ad_{X_0}W_0)\right]=-\Lambda_{\alpha}(X_0)gX_0.
 \end{equation}
Now we have to specify the factor $g_0:=\Lambda_{\alpha}(X_0)g$, that is a solution of the equation $(X_0-\alpha)g_0=0$. Indeed, since $\Lambda(X_0)g=(X_0-\alpha)\Lambda_{\alpha}(X_0)g$ then we get the former equation which implies $g_0=A(u_{1},u_{-1})e^{\alpha{u_0}}$, where $A(u_{1},u_{-1})$ is a nonzero quasi-polynomial on $u_1,u_{-1}$. The second formula of (\ref{lop}) is proved in a similar way.

We define a sequence of multiple commutators in such a way 
 \begin{equation*}
 R_1=\left[P_{\alpha},P_{\beta}\right],\quad R_2=\left[P_{\alpha},R_1\right],\quad...,\quad R_{k+1}=\left[P_{\alpha},R_k\right],\quad...
 \end{equation*}
Let us evaluate the commutator $\left[D_x, R_1 \right]$. Due to the Jacobi identity we have
\begin{eqnarray}  
\fl \left[D_x,R_1\right] = \left[ D_x, \left[P_{\alpha}, P_{\beta} \right]\right] = \left[ P_{\alpha}, \left[ D_x, P_{\beta}  \right]  \right] - \left[P_{\beta}, \left[ D_x, P_{\alpha} \right] \right] = \nonumber\\
= \left[P_{\alpha},B(u_1, u_{-1})e^{\beta{u_0}}X_0\right]-\left[P_{\beta},A(u_1, u_{-1})e^{\alpha{u_0}}X_0\right].	\label{eq40} 
\end{eqnarray}
By construction the vector fields $P_{\alpha}$, $P_{\beta}$ do not contain differentiation with respect to the variables $u_1, u_0, u_{-1}$, therefore  (\ref{eq40}) implies
\begin{equation}	\label{eq_41}
\left[ D_x, R_1 \right] = -B(u_1, u_{-1}) e^{\beta u_0} \left[ X_0, P_{\alpha} \right] + A(u_1, u_{-1}) e^{\alpha u_0} \left[ X_0, P_{\beta} \right].
\end{equation}
It remains to evaluate the commutators $\left[ X_0, P_{\alpha} \right]$ and $\left[X_0, P_{\beta}  \right]$. To this aim find their commutators with $D_x$:
\begin{equation*}
\left[D_x, \left[X_0, P_{\alpha}  \right]  \right] = \alpha A e^{\alpha u_0} P_{\alpha}, \qquad \left[D_x, \left[X_0, P_{\beta}  \right]  \right] = \beta B e^{\beta u_0} P_{\beta}.
\end{equation*} 
Now due to (\ref{lop}) we get 
\begin{equation*}
\left[ D_x, \left[X_0, P_{\alpha} \right] - \alpha P_{\alpha} \right] = 0, \qquad \left[ D_x, \left[X_0, P_{\beta} \right] - \beta P_{\beta} \right] = 0.
\end{equation*}
The last two equations imply in virtue of Lemma 1 the desired relations $\left[X_0, P_{\alpha} \right] = \alpha P_{\alpha}$, $\left[ X_0, P_{\beta} \right] = \beta P_{\beta}$. Thus (\ref{eq_41}) gives rise to
\begin{equation*}
\left[ D_x, R_1 \right] = -\alpha B e^{\beta u_0} P_{\alpha} + \beta A e^{\alpha u_0} P_{\beta}.
\end{equation*}
By the same way we find $\left[ X_0, R_1 \right] = (\alpha + \beta) R_1$ and then deduce the relation
\begin{equation*}
\left[ D_x, R_2 \right] = (\alpha + 2 \beta) A e^{\alpha u_0} R_1.
\end{equation*}
It can be proved by induction, that
 \begin{equation*}
 \left[D_x,R_m\right]=y_{m}R_{m-1}, \quad \left[X_0,R_{m-1}\right]=z_{m-1}{R_{m-1}}, \quad m \geq 2,
 \end{equation*}
 where $y_k$ and $z_k$ are solutions to the discrete equations
 \begin{equation}  \label{yz}
 y_{m+1}=y_m+Ae^{\alpha{u_0}}z_m, \quad z_m=z_{m-1}+\alpha, \quad m \geq 2
 \end{equation}
 with the following initial data
 \begin{equation}\label{bb}
 z_1=\alpha+\beta, \quad y_2=(\alpha+2\beta)Ae^{\alpha{u_0}}.
 \end{equation}
The problem (\ref{yz}), (\ref{bb}) is solved explicitly:
\begin{equation*}  \label{znyn}
z_n = \alpha n + \beta, \quad y_n = A e^{\alpha u_0}  (n + 1) \left( \frac{n}{2} \alpha + \beta \right).
\end{equation*}
Since the characteristic algebra is finite dimensional there exists a natural $N$ such that $R_{N+1}$ is linearly expressed through the previous members of the sequence:
 \begin{equation}\label{www}
 R_{N+1}={\lambda}_N{R_N}+{\lambda}_{N-1}R_{N-1}+...{\lambda}_1{R_1}+{\lambda}_{\alpha}P_{\alpha}+{\lambda}_{\beta}P_{\beta},
 \end{equation}
 where  the operators $R_{N},R_{N-1},...R_1,P_{\alpha},P_{\beta}$ are supposed to be linearly independent.  Applying the operator $ad_{D_x}$ to both sides of equation (\ref{www}), we find
 \begin{equation*}
 y_{N+1}R_N=D_x({\lambda}_N)R_N+{\lambda}_N y_N R_{N-1}+...
 \end{equation*}
 Collecting coefficients before $R_N$, we find the equation $D_x({\lambda}_N)=y_{N+1}$. We concentrate on this equation by representing it in an explicit form
$$\sum_{j}\frac{\partial \lambda_N}{\partial u_{j}}u_{j,x}+\sum_{j}\frac{\partial \lambda_N}{\partial u_{j,x}}u_{j,xx}+...=y_{N+1},$$ 
where $y_{N+1} = A e^{\alpha u_0}  (N + 2) \left( \frac{N+1}{2} \alpha + \beta \right)$.
Since the r.h.s. does not contain the variables $u_{j,x}$, $u_{j,xx}$,... we get immediately that $D_x({\lambda}_N)=0$.  Hence, this equation is satisfied only when ${\lambda}_N=const$ and $y_{N+1}=0$, or when $\frac{N+1}{2} \alpha + \beta = 0$. We can repeat all the reasoning by replacing ${\alpha}\leftrightarrow{\beta}$. Then we arrive at a similar relation with some natural $K$: $\frac{K+1}{2} \alpha + \beta = 0$. In other words the following system of equations
\begin{equation*}
(N + 1) \alpha + 2 \beta = 0, \quad (K + 1)\beta + 2 \alpha = 0
\end{equation*}
should have solution in natural $N$, $K$. Solving the system we get: $(N + 1) (K + 1) = 4$ when $\alpha\beta\neq0$. Note that if $\alpha\beta=0$ then both of the roots vanish. However that contradicts the requirement $\alpha\neq \beta$. Thus we have either $N = K = 1$ or $K = 0$, $N = 3$. In the first case $\alpha = -\beta$, in the second case $\beta = - 2 \alpha$. This completes the proof of Lemma 5.  

These two exceptional cases are studied in the following theorem.

{\bf Theorem 2.} \textit{If the polynomial (\ref{Lambda_00}) has two different roots $\alpha$ and $\beta=-2\alpha$ (or $\alpha$ and $\beta=-\alpha$) then the characteristic Lie-Rinehart algebra corresponding to the reduced system (\ref{finite_sys}) is of infinite dimension.}

In other words Theorem~2 claimes that in these two cases the reduced system is not integrable in the sense of Darboux. Theorem~2 is proved in Appendix.

\subsection{Investigation of multiple roots}

Now let's study the problem of the multiplicity of the roots of the polynomial $\Lambda(\lambda)$ defined by (\ref{Lambda_00}). 

{\bf Lemma 6.} \textit{Polynomial $\Lambda(\lambda)$ does not have any multiple non-zero root.}

{\bf Proof.} Suppose that $\lambda=\alpha$ is a multiple non-zero root of the  polynomial $\Lambda(\lambda)$. Define new  polynomials $\Lambda_1(\lambda)=\frac{1}{\lambda-\alpha}\Lambda(\lambda)$ and $\Lambda_2(\lambda)=\frac{1}{(\lambda-\alpha)^2}\Lambda(\lambda)$.
    Then we consider the equation ${\Lambda}_1(X)g(u_{1},u_0,u_{-1})=y$. Evidently $(X_0-\alpha)y=0$, therefore by solving this equation one can find $y=A(u_1,u_{-1})e^{{\alpha}u_0}$. Similarly we put $\Lambda_2(X)g(u_{1}, u_0, u_{-1})=z$ and then find that $(X_0-\alpha)z={\Lambda}_1(X)g=y$, which implies that $z=e^{{\alpha}u_0}\left(A(u_{1},u_{-1})u_0+B(u_{1},u_{-1})\right)$, where $B=B(u_1,u_{-1})$ is a function.
    
    Due to the formula (\ref{ss}) we can obtain that the operators $P={\Lambda}_{1}(ad_{X_0}W_0)$, 
	$T={\Lambda}_{2}(ad_{X_0}W_0) $ satisfy the following commutativity conditions
	\begin{eqnarray*}
    && \left[D_x,P\right]=A(u_{1},u_{-1})e^{{\alpha}u_0}X_0,\label{op1}\\
		&& 	\left[D_x,T\right]=e^{{\alpha}u_0}\left(A(u_{1},u_{-1})u_0+B(u_{1},u_{-1})\right)X_0. \label{op2}
   \end{eqnarray*}
Let us construct a sequence of the operators due to the formulas:
   \[
    K_1=\left[P,T\right], \quad K_2=\left[P,K_1\right], \quad ..., \quad K_{m+1}=\left[P,K_m\right],\quad ...
 \]
    It is easily checked that $\left[X_0,P\right]={\alpha}P$, $\left[X_0,T\right]={\alpha}T+P$. Indeed, let us check the first of these formulas. Evidently we have $\left[D_x,\left[X_0,P\right]\right]=\left[X_0,Ae^{{\alpha}u_0}X_0\right]={\alpha}Ae^{{\alpha}u_0}X_0$. Therefore, $\left[D_x,\left[X_0,P\right]-{\alpha}P\right]=0.$ By virtue of Lemma 1 one obtains the  formula desired. In a similar way we prove that 
    \begin{eqnarray*}
    &&\left[D_x,K_1\right]=e^{{\alpha}u}({\alpha}Au+{\alpha}B-A)P-e^{{\alpha}u_0}{\alpha}AT, \\
			&&\left[D_x,K_2\right]=3{\alpha}Ae^{{\alpha}u_0}K_1, \\
			&&\left[X_0,K_1\right]=2{\alpha}K_1, \quad \left[X_0,K_2\right]=3{\alpha}K_2. 
    \end{eqnarray*} 
    It can be proved by induction that for any $m{\geq}2$
    \begin{equation*}
    \left[D_x,K_m\right]=\frac{\alpha}{2}(m+1)mAe^{{\alpha}u_0}K_{m-1}.
        \end{equation*}
        Since the Lie-Rinehart algebra generated by $P,T$ is supposed to be of a finite dimension then there is an integer $M$ such that
        \begin{equation}\label{ser}
        K_{M+1}=a_{M}K_{M}+a_{M-1}K_{M-1}+...+a_1{K_1}+{b_1}P+{b_2}T,
        \end{equation}
        where $a_j,b_j$ are some functions depending on the dynamical variables $u_j,u_{jx},u_{jxx},...$ and the operators $K_M,K_{M-1},...K_1,P,T$ are linearly independent. By applying the operator $ad_{D_x}$ to equation (\ref{ser}) one gets for $M > 0$
        \[
         \frac{1}{2}{\alpha}(M+2)(M+1)Ae^{{\alpha}u_0}K_M=D_x(a_M)K_M+...,
         \]
          where the tail contains the summands with  $K_{M-1},K_{M-2},...$. Thus  the last equation implies 
        \begin{equation}\label{cont}
        D_x(a_M(\bar{u},\bar{u}_x,\bar{u}_{xx},...))=\frac{\alpha}{2}(M+2)(M+1)A(u_{1},u_{-1})e^{{\alpha}u_0}.
        \end{equation}
        Equation (\ref{cont}) yields $D_x(a_M)=0$ and $\alpha=0$. The latter contradicts the assumption $\alpha\neq0$.
				
				The case $M=0$ i.e. $K_1={b_1}P+{b_2}T$ should be investigated separately.        
        Here application of the operator $ad_{D_x}$ yields
        \begin{eqnarray*}
				&&e^{{\alpha}u_0}({\alpha}Au+{\alpha}B-A)P-{\alpha}e^{{\alpha}u_0}AT=\\
				&&=D_x(b_1)P+D_x(b_2)T+{b_1}Ae^{{\alpha}u_0}X_0+{b_2}e^{{\alpha}u_0}(Au_0+B)X_0.
				\end{eqnarray*}
        Comparison of the coefficients before the operators $X_0,P,T$ shows that this equation is contradictory. This completes the proof of Lemma~6.

{\bf Lemma 7.} \textit{At the point $\lambda = 0$ polynomial $\Lambda(\lambda)$ defined in (\ref{Lambda_00}) might have only a simple root.}

{\bf Proof.} Assume that $\alpha=0$ is a root the characteristic polynomial ${\Lambda}(\lambda)$ of the multiplicity $k$. Then then due to Theorem 2 we have ${\Lambda}(\lambda)={\lambda}^k$.  
Let us construct operators:
\begin{eqnarray*}
&&P_1={\Lambda}_1(ad_{X_0}{W_0})={ad^{k-1}_{X_0}}{W_0},\\
&&P_2={\Lambda}_2(ad_{X_0}{W_0})={ad^{k-2}_{X_0}}{W_0}.
\end{eqnarray*}
It can be proved that they satisfy relations:
 \begin{eqnarray*}
 &&\left[D_x,P_1\right]=-A(u_{1},u_{-1})X_0,\\
 &&\left[D_x,P_2\right]=-\left(A(u_{1},u_{-1})u_0+B(u_{1},u_{-1})\right)X_0.
 \end{eqnarray*}
 
 Let us prove that the Lie-Rinehart algebra generated by $P_1$ and $P_2$, where the coefficient $A(u_{1},u_{-1})$ doesn't vanish identically, is of infinite dimension. Define a sequence of the multiple commutators in such a way 
\[
P_1, \quad P_2, \quad P_3=\left[P_2, P_1\right], \quad ..., \quad P_m=\left[P_2, P_{m-1}\right], \quad ...
\]
  It can be proved by induction on $m$ that 
  \[
  \left[D_x,P_m\right]=A(u_{1},u_{-1})P_{m-1}.
 \]
 If the Lie-Rinehart algebra is of a finite dimension then there exists a natural $M$ such that
\[
P_{M+1}={\mu}_M{P_M}+{\mu}_{M-1}{P}_{M-1}+...+{\mu}_1P_1.
\]
By applying the operator $D_x$ to both sides of the last equality, we obtain:
\begin{eqnarray*}
A(u_{1},u_{-1})P_{M}&&=D_x({\mu}_M)P_M+{\mu}_MA(u_{1},u_{-1})P_{M}+...+\\
&&+D_x({\mu}_1)P_1+{\mu}_1A(u_{1},u_{-1})P_{1}.
\end{eqnarray*}
 By comparing the coefficients before $P_M$ we get:
 \begin{equation}\label{ool}
 A(u_{1},u_{-1})=D_x({\mu}_M).
 \end{equation}
 Since ${\mu}_M={\mu}_M(\bar{u},\bar{u}_x,\bar{u}_{xx},...)$ depends on a set of the dynamical variable while $A$ depends only on $u_{1}$ and $u_{-1}$   equality (\ref{ool}) fails to be true unless $A(u_{1},u_{-1})=0$ that contradicts to our assumption. Therefore the Lie-Rinehart algebra generated by the operators $P_1$ and $P_2$ is of infinite dimension. Lemma~7 is proved.

Thus, summarizing the statements of Lemmas~5--7, we conclude  that the polynomial $\Lambda(\lambda)$ defined by (\ref{Lambda_00}) might  have only one root and this root is simple. Therefore, equation (\ref{Lambda_0}) has the form $(X_0 - \alpha)f_{u_1 u_{-1}} = 0$, where $\alpha$ is a constant. Thus, we have that the function $f_{u_1 u_{-1}}$ has the following form:
\begin{equation*}
f_{u_1 u_{-1}} = Q(u_{1}, u_{-1}) e^{\alpha u_0},
\end{equation*}
where $Q(u_1 u_{-1})$ is a function being a quasi-polynomial with respect to any of its arguments $u_{1}$, $u_{-1}$.

Now, let us repeat the reasoning above by changing the operator $X_0$ by $X_1$. For this purpose, we construct a sequence as follows
\begin{equation*}
H_0 = W_0, \quad H_1 =\left[X_1, H_0\right], \quad H_2 = \left[X_1, H_1\right], \ldots, \quad H_{k+1} = \left[X_1, H_k\right], \ldots
\end{equation*}
Elements of the sequence satisfy the relations: 
\begin{equation*} 
\left[ D_x, H_k  \right] = -X^{k}_1(f_{u_1 u_{-1}})X_0 = -e^{\alpha u_0}X^{k}_1(Q_{u_1 u_{-1}})X_0 .
\end{equation*}
Since the characteristic algebra is finite-dimensional there exists a natural $K$ such that $H_{K+1}$ is linearly expressed by the previous members:
\begin{equation*}
H_{K+1} + \lambda_{K} H_K + \cdots + \lambda_1 H_1 + \lambda_0 H_0 = 0,
\end{equation*}
where the operators $H_0, H_1, \ldots, H_k$ are linearly independent. 
We apply the operator $ad_{D_x}$ to the obtained equation and get relations $D_x(\lambda_j)=0$ for $j=0,1,...K$ and also a relation $\Omega(X_1)Q_{u_1u_{-1}}=0$, where  
\begin{equation}\label{3*}
\Omega(\lambda)=\lambda^{K+1}  + \lambda_{K} \lambda^{K} + \cdots  + \lambda_0  
\end{equation}
is a quasi-polynomial with constant coefficients. 

Now we investigate the characteristic polynomial (\ref{3*}) by using the reasonings we applied above to the characteristic polynomial (\ref{Lambda_00}). As a result we prove that $Q_{u_{1} u_{-1}} = \Phi(u_{-1}) e^{\beta u_1}$ and $f_{u_{1} u_{-1}} = \Phi(u_{-1}) e^{\alpha u_0 + \beta u_1}$. Here $\Phi(u_{-1})$ is a quasi-polynomial.

Finally, we repeat this reasoning, replacing $X_{1}$ by $X_{-1}$ and  prove the following statement:

{\bf Theorem 3.} \textit{If lattice (\ref{eq_main}) is integrable in the sense of Definition 1 then the function $f(u_{n+1}, u_n, u_{n-1})$ satisfies the following equation:}
 \begin{equation}\label{fu1um1}
 f_{u_{n+1}, u_{n-1}}=Ce^{{\alpha}u_n+{\beta}u_{n+1}+{\gamma}u_{n-1}},
 \end{equation}
 \textit{where $C,{\alpha},{\beta},{\gamma}$ are constant.}

\section{The necessary integrability conditions}


{\bf Theorem 4.} 
\textit{If $\alpha = 0$, $C \neq 0$ in (\ref{fu1um1}) then $\beta = 0$ and $\gamma = 0$. }
\textit{If $\alpha \neq 0$, $C \neq 0$ then $\beta = -\frac{\alpha}{2}m$, $\gamma = -\frac{\alpha}{2}k$, where $m$, $k$ are nonnegative integers.}

{\bf Proof.} Suppose that $C \neq 0$ then a minimal order operator (\ref{Lambda_0}) which annulates the function $g = f_{u_1 u_{-1}}$ has the form $\Lambda(X_0) = X_0 - \alpha$. Let us construct the operator $P_0 = \Lambda(ad_{X_0}W_0) / C(\lambda - \alpha)$ and the operator $P_1 = D_nP_0D_n^{-1}$, where $D_n$ stands for the shift operator acting as $D_ny(n)=y(n+1)$.  Thus we have the operators $P_0, P_1\in{L}$ of the form
\begin{equation*}
P_j=\sum_{k}{a_{jk}(1)\frac{\partial}{\partial{u_{kx}}}}+{a_{jk}(2)\frac{\partial}{\partial{u_{kxx}}}}+{a_{jk}(3)\frac{\partial}{\partial{u_{kxxx}}}}+...
\end{equation*}
Due to (\ref{fu1um1}) and Lemma 4 these operators satisfy the following commutativity relations
\begin{equation}\label{bvc}
\left[D_x,P_0\right]=e^{\omega}X_0, \quad \left[D_x,P_1\right]=e^{\omega_1}{X_1},
\end{equation}
where $\omega={\alpha}u_0+{\beta}u_1+{\gamma}u_{-1},{\omega}_1={\alpha}u_1+{\beta}u_2+{\gamma}u_0$. The first relation in (\ref{bvc}) is easily proved by the same way as formula (\ref{lop}). The second relation follows from the first one by applying the conjugation transformation $X\rightarrow D_nXD_n^{-1}$.
 
 One can easily verify that
\begin{equation*}
\left[X_0,P_0\right]={\alpha}P_0, \quad \left[X_0,P_1\right]={\gamma}P_1, \quad \left[X_1,P_0\right]={\beta}P_0, \quad \left[X_1,P_1\right]={\alpha}P_1.
\end{equation*}
 Let us  construct a sequence of the multiple commutators as follows
 \begin{equation*}
 K_1=\left[P_0,P_1\right], \quad K_2=\left[P_0,K_1\right],..., \quad K_m=\left[P_0,K_{m-1}\right], ...
 \end{equation*}
 It is checked by direct calculation that
 \begin{eqnarray*}
 &&\left[D_x,K_1\right]=-{\beta}e^{{\omega}_1}P_0+{\gamma}e^{\omega}P_1, \quad \left[X_0,K_1\right]=(\alpha+\gamma)K_1,\\
 &&\left[D_x,K_2\right]=(\alpha+2\gamma)e^{\omega}K_1, \quad \left[X_0,K_2\right]=(2\alpha+\gamma)K_2.
 \end{eqnarray*}
 By induction we can prove that for $n\geq{2}$
 \begin{equation}\label{tty}
 \left[D_x,K_n\right]=e^{\omega}{\pi}_n K_{n-1},\quad \left[X_0,K_n\right]=y_n{K_n}
 \end{equation}
 with $\pi_{n}=\frac{\alpha}{2} n^2+(\gamma-\frac{\alpha}{2})n$, $y_n=\alpha n+\gamma$. 
 Since the characteristic algebra is of a finite dimension then there exists a natural $N$ such that 
 \begin{equation}\label{pol}
 K_{N+1}={\lambda}_N{K_N}+...+{\lambda}_1{K_1}+{\mu}_0{P_0}+{\mu}_1{P_1},
 \end{equation}
 where the operators $P_0$, $P_1$, $K_1$,..., $K_N$ are linearly independent. By applying to (\ref{pol}) the operator $ad_{D_x}$ one find  due to (\ref{tty}) that
 \begin{equation}\label{vvc}
 e^{\omega}{\pi}_{N+1}K_N={D_x}(\lambda_N){K_N}+{\lambda}_N{e^{\omega}}{\pi}_N{K_{N-1}}+...
 \end{equation}
 We  compare the coefficients before $K_N$ in (\ref{vvc}) and we get ${\lambda}_N=0$, ${\pi}_{N+1}=0$. Or the same
\begin{equation} \label{aNg}
\frac{\alpha}{2} N + \gamma = 0.
\end{equation}
From this formula it follows that if $\alpha = 0$ then $\gamma = 0$. Similarly one can prove that if $\alpha = 0$ then $\beta = 0$.

The case $N=0$ is never realized. Indeed supposing $K_1={\mu}_0{P_0}+{\mu}_1{P_1}$ one obtains a contradictory equation
 \begin{equation*}
  -{\beta}e^{{\omega}_1}P_0+{\gamma}e^{\omega}P_1 =D_x({\mu}_0)P_0+D_x({\mu}_1)P_1+{\mu}_0e^{\omega}X_0+{\mu}_1e^{{\omega}_1}X_1
 \end{equation*}
 unless $\beta=0$, $\gamma=0$. 

 Thus it follows from (\ref{aNg}) that $\gamma = -\frac{\alpha}{2}N$. Formula $\beta = -\frac{\alpha}{2}m$ is proved in a similar way.  Moreover, we see that if $\gamma = 0$ then $\beta = 0$. And similarly we obtain that if $\beta = 0$ then $\gamma = 0$.   In other words, if $\beta \gamma \neq 0$ then $f_{u_1 u_{-1}}$ has the following form
 \begin{equation*}  
  f_{u_1 u_{-1}}=Ce^{{\alpha}u_0-{\frac{{\alpha}m}{2}}u_1-{\frac{{\alpha}k}{2}}u_{-1}},
 \end{equation*}
where $C \neq 0$, $\alpha \neq 0$. If $\beta = 0$ or $\gamma = 0$ then both $\beta = \gamma = 0$ and 
\begin{equation*} 	
  f_{u_1 u_{-1}}=Ce^{{\alpha}u_0},
 \end{equation*}
where $\alpha \neq 0$, $C \neq 0$.
  Theorem 4 is proved.

The main result is given in 

{\bf Theorem 5.} \textit{Lattice (\ref{eq_main}) which is integrable in the sense of Definition 1,  can be reduced by suitable rescalings to one of the following forms:}
\begin{eqnarray*}	
&& u_{n,xy} =  e^{\alpha u_n - \frac{\alpha}{2}m u_{n+1} - \frac{\alpha}{2} k u_{n-1}} + a(u_{n+1}, u_n) + b(u_n, u_{n-1}), \\
&& u_{n,xy} =  e^{\alpha u_n} u_{n+1} u_{n-1} + a(u_{n+1}, u_n) + b(u_n, u_{n-1}), \\
&& u_{n,xy} =  u_{n+1} u_{n-1} + a(u_{n+1}, u_n) + b(u_n, u_{n-1}), \\
&& u_{n,xy} =  a(u_{n+1}, u_n) + b(u_n, u_{n-1}); 
\end{eqnarray*}
\textit{here  $\alpha \neq 0$ and $m$, $k$ are positive integers.}





Theorem 5 straightforwardly follows from Theorems 3, 4.

\section{Appendix}

Here we give a complete proof of Theorem 2. We consider the cases $\beta = -2 \alpha$ and $\beta = - \alpha$ separately. Our proof uses the scheme applied earlier in \cite{ZhMurt}, \cite{Sakieva}.

\subsection{The case $\beta = -2 \alpha$}

In this subsection we will prove that if polynomial $\Lambda(\lambda)$ has two different nonzero roots $\alpha$ and $\beta = -2 \alpha$ then the Lie-Rinehart algebra $L$ generated by the operators $X_0$ and $W_0$ is of an infinite dimension.

First we introduce two polynomials according to the rule
\begin{equation*} 
{\Lambda}_{\alpha}(\lambda)=\frac{\Lambda(\lambda)}{\lambda-\alpha}, \quad {\Lambda}_{\beta}(\lambda)=\frac{\Lambda(\lambda)}{\lambda+ 2 \alpha}.
\end{equation*}
Then we construct two operators $P_{\alpha}, P_{\beta} \in L$:
\begin{equation*}
P_{\alpha}={\Lambda}_{\alpha}(ad_{X_0}W_0), \quad P_{\beta}={\Lambda}_{\beta}(ad_{X_0}W_0)
\end{equation*}
and concentrate on the Lie-Rinehart algebra $L_{1} \subset L$ being a subalgebra of $L$ generated by the operators $P_{\alpha}, P_{\beta}$. By construction these operators satisfy the following commutativity relations
\begin{equation*}	
\left[ D_x, P_{\alpha}  \right] = A(u_{1}, u_{-1}) e^{\alpha u_0} X_0, \quad \left[ D_x, P_{\beta} \right] = B(u_{1}, u_{-1}) e^{-2 \alpha u_0} X_0,
\end{equation*}
where $A(u_{1}, u_{-1})$, $B(u_{1}, u_{-1})$ are quasi-polynomials in $u_{1}$, $u_{-1}$. We assume that $(u_{-1}, u_1) \in D$, where $D$ is a domain in $\mathbb{C}^2$, where both $A$, $B$ do not vanish.

Let us consider the operators:
\begin{equation}  \label{X1X2}
Y_1 = P_{\alpha} + P_{\beta}, \quad Y_2 = \frac{\partial}{\partial u_0}.
\end{equation}

Assume that $L_i$ stands for the linear space spanned by all possible commutators of the operators $Y_1$ and $Y_2$ of the length less or equal to $i-1$, where $i = 2, 3, \ldots$. We emphasize that the linear combination in this space is taken with coefficients being the functions depending on a finite number of the variables  $\bar u,\bar u_x,\bar u_{xx},\dots$. Thus $L_2 = \left\{ Y_1, Y_2 \right\}$ is the linear span of $Y_1$ and $Y_2$, $\mathrm{dim}\, L_2 = 2$. Similarly $L_3$ is the linear envelop of the vector $Y_1$, $Y_2$ and $Y_3 = \left[ Y_1, Y_2 \right]$, i.e. $L_3 = \left\{Y_1, Y_2, Y_3 \right\}$. Therefore, $L_4 = \left\{Y_1, Y_2, Y_3, \left[Y_1, Y_3 \right], \left[Y_2, Y_3 \right] \right\}$ etc.

Let us denote  $\delta(i) = \mathrm{dim}\, L_i - \mathrm{dim}\, L_{i-1}$. We also will use the following notations for multiple commutators:
\begin{equation*}
Y_{i_1,\ldots, i_n} = ad_{Y_{i_1}}\ldots ad_{Y_{i_{n-1}}} Y_{i_n}, \quad \mathrm{where} \quad ad_Y W= \left[ Y, W \right].
\end{equation*}

{\bf Lemma 8.} \textit{Assume that polynomial $\Lambda(\lambda)$ has two different nonzero roots $\alpha$ and $-2 \alpha$. Then the following formulas are true:}
\begin{equation*}
\delta(i) = 2, \quad i= 6n + 2, \quad i=6n+4, \quad n = 1,2, \ldots,
\end{equation*}
\begin{equation*}
\delta(i) = 1, \quad i = 6n-1, \quad i = 6n, \quad i = 6n+1, \quad i = 6n+3, \quad n=1,2,\ldots,
\end{equation*}
\begin{eqnarray*}
L_{6n+2} = L_{6n+1} \oplus \left\{ Y_{1 \ldots 121}, Y_{ 2 1 \ldots 121}  \right\},\\
L_{6n+4} = L_{6n+3} \oplus \left\{  Y_{1 \ldots 121}, Y_{ 2 1 \ldots 121}  \right\},\\
L_{6n-1} = L_{6n-2} \oplus \left\{ Y_{1 \ldots 121}\right\},\\
L_{6n} = L_{6n - 1} \oplus \left\{Y_{1 \ldots 121}\right\},\\
L_{6n+1} = L_{6n} \oplus \left\{Y_{1 \ldots 121}\right\},\\
L_{6n+3} = L_{6n+2} \oplus \left\{Y_{1 \ldots 121}\right\}.
\end{eqnarray*}

{\bf Proof.} We introduced the operators $Y_1$, $Y_2$ by formulas (\ref{X1X2}). The following commutation relations are true for these operators:
\begin{equation}  \label{DxX1X2}
\left[ D_x, Y_1  \right] = \left( A e^{\alpha u_0} + B e^{-2 \alpha u_0} \right) Y_2,  \quad \left[ D_x, Y_2 \right] = 0.
\end{equation}
We introduce the operator of length 2: $Y_3 = \left[ Y_2, Y_1 \right] = Y_{21}$. Then using the Jacobi identity and formulas (\ref{DxX1X2}), we get
\begin{equation}	\label{DxX3}
\left[ D_x, Y_3 \right] = \alpha \left( A e^{\alpha u_0} - 2 B e^{-2 \alpha u_0}  \right) Y_2.
\end{equation}
If we assume that $Y_3$ is linearly expressed by $Y_1$ and $Y_2$, i.e.
\begin{equation}  \label{linX3}
Y_3 = \lambda_1 Y_1 + \lambda_2 Y_2,
\end{equation}
then we get a contradiction. Indeed by commuting both sides of (\ref{linX3}) with $D_x$ and then simplifying due to (\ref{DxX1X2}), (\ref{DxX3}) we obtain
\begin{equation*}
\fl \alpha \left( A e^{\alpha u_0} - 2 B e^{-2 \alpha u_0} \right) Y_2 = D_x(\lambda_1) Y_1 + \lambda_1 \left( A e^{\alpha u_0} + B e^{-2 \alpha u_0} \right)Y_2 + D_x(\lambda_2) Y_2.
\end{equation*}
Comparing the coefficients before independent operators $Y_1$, $Y_2$, we get: $D_x(\lambda_1) = 0$ and
\begin{equation*}
\alpha \left( A e^{\alpha u_0} - 2 B e^{-2 \alpha u_0} \right) = \lambda_1 \left( A e^{\alpha u_0} + B e^{-2 \alpha u_0}  \right) + D_x(\lambda_2).
\end{equation*}
The last equality implies that $D_x(\lambda_2) = 0$  and $\lambda_1 - \alpha = 0$, $\lambda_1 + 2 \alpha = 0$. Obviously a pair of these equations is inconsistent because  $\alpha \neq 0$.

We introduce the commutators of length 3: $Y_4 = \left[Y_1, Y_3 \right]$ and $\bar{Y}_4 = \left[Y_2, Y_3 \right]$. Then
\begin{eqnarray}
\left[D_x, \bar{Y}_4  \right] = \alpha^2 \left( A e^{\alpha u_0} + 4B e^{-2 \alpha u_0} \right)Y_2, \label{DxX4}\\
\left[D_x, Y_4 \right] = -\alpha \left( 2 A e^{\alpha u_0} - B e^{-2 \alpha u_0} \right)Y_3 + 2 \alpha^2 \left( A e^{\alpha u_0} + B e^{-2 \alpha u_0} \right)Y_1. \nonumber
\end{eqnarray}
One can see that $\left[ D_x, \bar{Y}_4 \right] = 2 \alpha^2 \left[ D_x, Y_1\right] - \alpha \left[D_x, Y_3\right] = \left[D_x, 2 \alpha^2 Y_1 - \alpha Y_3\right]$. 
Due to Lemma 1 this equality implies that $\bar{Y}_4 = 2 \alpha^2 Y_1 - \alpha Y_3$. The operator $Y_4 = Y_{121}$ is not linearly expressed through the operators of lower order. Thus, we have $L_4 = \left\{ Y_1, Y_2, Y_3, Y_4 \right\}$.

We introduce the commutators of lenght 4: $Y_5 = \left[Y_1, Y_4 \right]$ and $\bar{Y}_5 = \left[Y_2, Y_4 \right]$. Using the Jacobi identity and formulas (\ref{DxX1X2}), (\ref{DxX4}), we find
\begin{eqnarray*}
\fl \left[D_x, \bar{Y}_5  \right] = \alpha^2 \left( 2A e^{\alpha u_0} - B e^{-2 \alpha u_0} \right)Y_3 - 2 \alpha^3 \left( A e^{\alpha u_0} + B e^{-2 \alpha u_0} \right)Y_1 = \left[  D_x, -\alpha Y_4 \right],\\
\fl \left[ D_x, Y_5 \right] = -\alpha \left( 2 A e^{\alpha u_0} - B e^{-2 \alpha u_0} \right) Y_4 - \left( A e^{\alpha u_0} + B e^{-2 \alpha u_0} \right) \left[ Y_4, Y_2\right]=    - 3 \alpha A e^{\alpha u_0} Y_4.
\end{eqnarray*}
Due to Lemma 1, we conclude that $\bar{Y}_5 = -\alpha Y_4$. The operator $Y_5 = Y_{1121}$ is not linearly expressed through the commutators of lower order. Thus, we have $L_5 = \left\{Y_1, Y_2, Y_3, Y_4, Y_5 \right\}$.

Let us consider the commutators of the length 5:
\begin{equation*}
Y_6 = \left[Y_1, Y_5\right], \quad \bar{Y}_6 = \left[Y_2, Y_5\right], \quad \left[Y_3, Y_4\right]
\end{equation*}
The following formulas are true:
\begin{equation*}
\fl \left[ D_x, \bar{Y}_6 \right] = 0, \quad \left[ D_x, \left[Y_3, Y_4  \right]\right] = -3 \alpha^2 A e^{\alpha u_0} Y_4 = \left[D_x, \alpha Y_5 \right], \quad \left[ D_x, Y_6 \right] = - 3 \alpha A e^{\alpha u_0} Y_5.
\end{equation*}
Using Lemma 1, we conclude that $\left[Y_3, Y_4 \right] = \alpha Y_5$. The operator $Y_6 = Y_{11121}$ is not linearly expressed by the operators of lower order. So, we have $L_6 = \left\{ Y_1, Y_2, Y_3, Y_4, Y_5, Y_6 \right\}$.

Now we introduce the operators of the length 6:
\begin{equation*}
Y_7 = \left[Y_1, Y_6\right] = Y_{111121}, \quad \bar{Y}_7 = \left[Y_2, Y_6\right] = Y_{211121}, \quad \left[Y_3, Y_5\right].
\end{equation*}
One can prove that the following formulas are true:
\begin{equation*}
\bar{Y}_7 = \alpha Y_6, \quad \left[Y_3, Y_5\right] = \alpha Y_6, \quad \left[ D_x, Y_7 \right] = \alpha \left( -2 A e^{\alpha u_0} + B e^{-2 \alpha u_0} \right) Y_6.
\end{equation*}
The operator $Y_7 = Y_{111121}$ is not linearly expressed by the operators of lower order and $L_7 = \left\{ Y_1, Y_2, Y_3, Y_4, Y_5, Y_6, Y_7 \right\}$.

Then we consider the operators of the length 7:
\begin{equation*}
Y_8 = \left[ Y_1, Y_7\right], \quad \bar{Y}_8 = \left[ Y_2, Y_7 \right], \quad \left[ Y_3, Y_6 \right], \quad \left[ Y_4, Y_5 \right].
\end{equation*}
One can prove that
\begin{eqnarray*}
\left[D_x, \bar{Y}_8  \right] = \alpha^2 \left( -4 A e^{\alpha u_0} - B e^{-2 \alpha u_0} \right)Y_6,\\
\left[ D_x, Y_8 \right] = \alpha \left( -2 A e^{\alpha u_0} + B e^{-2 \alpha u_0} \right)Y_7 + \left( A e^{\alpha u_0} + B e^{-2 \alpha u_0} \right) \bar{Y}_8.
\end{eqnarray*}
The operators $Y_8$, $\bar{Y}_8$ is not expressed through the operators of lower order. It is not difficult to show that $\left[Y_3, Y_6 \right] = \bar{Y}_8 - \alpha Y_7$, $\left[ Y_4, Y_5 \right] = 2 \alpha Y_7 - \bar{Y}_8$. Thus we have that the space $L_8$ is obtained from $L_7$ by adding two linearly independent elements $Y_8 = Y_{1111121}$ and $\bar{Y}_8 = Y_{2111121}$, i.e. $L_8  =\left\{Y_1, Y_2, Y_3, Y_4, Y_5, Y_6, Y_7, Y_8, \bar{Y}_8 \right\}$.

Now let us introduce the operators of the length 8:
\begin{equation*}
Y_9 = \left[Y_1, Y_8  \right], \quad \bar{Y}_9 = \left[Y_2, Y_8\right], \quad \left[ Y_1, \bar{Y}_8 \right], \quad \left[Y_2, \bar{Y}_8 \right], \quad \left[ Y_3, Y_7 \right], \quad \left[Y_4, Y_6 \right].
\end{equation*}
One can show that 
\begin{eqnarray*}
\left[Y_4, Y_6 \right] = \alpha Y_8, \quad \left[Y_3, Y_7 \right] = -\alpha Y_8, \quad \left[ Y_2, \bar{Y}_8 \right] = 2 \alpha^2 Y_7 + \alpha \bar{Y}_8,\\
\left[ Y_1, \bar{Y}_8 \right] = \alpha Y_8, \quad \bar{Y}_9 = 0,\\
\left[D_x, Y_9  \right] = \alpha \left( -A e^{\alpha u_0} + 2 B e^{-2 \alpha u_0} \right)Y_8.
\end{eqnarray*}
So we see that $Y_9 = Y_{11111121}$ is not linearly expressed through the operators of lower order and $L_9 = L_8 \oplus {Y_9}$.

The operators of length 9 are constructed by the following way:
\begin{equation*}
\fl Y_{10} = \left[Y_1, Y_9\right], \quad \bar{Y}_{10} = \left[Y_2, Y_9\right], \quad \left[Y_3, \bar{Y}_8\right], \left[ Y_3, Y_8 \right], \quad \left[Y_4, Y_7 \right], \quad \left[Y_5, Y_6\right].
\end{equation*}
For these operators the relations hold:
\begin{eqnarray*}
\left[ Y_3, \bar{Y}_8 \right] = -3 \alpha^2 Y_8, \quad \left[Y_3, Y_8  \right] = \bar{Y}_{10}, \quad \left[Y_4, Y_7\right] = -\alpha Y_9 - \bar{Y}_{10}, \\
 \left[Y_5, Y_6\right] = 2 \alpha Y_9 + \bar{Y}_{10}, \\
\left[D_x, Y_{10} \right] = \alpha \left( -A e^{\alpha u_0} + 2 B e^{-2 \alpha u_0}\right)Y_9 + \left( A e^{\alpha u_0} + B e^{-2 \alpha u_0} \right) \bar{Y}_{10},\\
\left[ D_x, \bar{Y}_{10} \right] = \alpha^2 \left( -A e^{\alpha u_0} - 4 B e^{-2 \alpha u_0} \right) Y_8.
\end{eqnarray*}
The operators $Y_{10} = Y_{111111121}$ and $\bar{Y}_{10} = X_{211111121}$ are not linearly expressed through the operators of  lower order, $L_{10} = L_9 \oplus \left\{Y_{10}, \bar{Y}_{10} \right\}$. 

Now let us introduce the notation: $Y_n = \left[Y_1, Y_{n-1} \right]$, $\bar{Y}_n = \left[Y_2, Y_{n-1} \right]$. 
We prove Lemma 7 by induction. Assume that for $i = n - 1$ the following formulas are true:
\begin{eqnarray}
\fl \left[D_x, Y_{6(n-1)-1} \right]= -\alpha \left( 2 A e^{\alpha u_0} - B e^{-2 \alpha u_0} \right) Y_{6(n-1)-2} +\label{Ind_01}\\
 \left( A e^{\alpha u_0} + B e^{-2 \alpha u_0} \left[Y_2, Y_{6(n-1)-2}\right] \right),\nonumber\\
\fl \left[ D_x, Y_{6(n-1)} \right] = -3 \alpha A e^{\alpha u_0} Y_{6(n-1)-1} + \left( A e^{\alpha u_0} + B e^{-2 \alpha u_0} \right) \left[Y_2, Y_{6(n-1)-1}  \right],\\
\fl \left[D_x, Y_{6(n-1)+1}  \right] = -3 \alpha A e^{\alpha u_0}Y_{6(n-1)} + \left( A e^{\alpha u_0} + B e^{-2 \alpha u_0} \right) \left[Y_2, Y_{6(n-1)}  \right],\\
\fl \left[D_x, Y_{6(n-1)+2}  \right] = \alpha \left( - 2 A e^{\alpha u_0} + B e^{-2 \alpha u_0}  \right) Y_{6(n-1)+1} + \nonumber\\
+ \left( A e^{\alpha u_0} + B e^{-2 \alpha u_0} \right) \left[Y_2, Y_{6(n-1)+1}  \right],\\
\fl \left[D_x, Y_{6(n-1)+3}  \right] = \alpha \left( -  A e^{\alpha u_0} + 2 B e^{-2 \alpha u_0}  \right) Y_{6(n-1)+2} +\nonumber \\
 +\left( A e^{\alpha u_0} + B e^{-2 \alpha u_0} \right) \left[Y_2, Y_{6(n-1)+2}  \right],\\
\fl \left[D_x, Y_{6(n-1)+4}  \right] = \alpha \left( -  A e^{\alpha u_0} + 2 B e^{-2 \alpha u_0}  \right) Y_{6(n-1)+3} +\nonumber\\
+ \left( A e^{\alpha u_0} + B e^{-2 \alpha u_0} \right) \left[Y_2, Y_{6(n-1)+3}  \right],\\
\fl \bar{Y}_{6(n-1)} = \left[Y_2, Y_{6(n-1)-1} \right] = 0,\\
\fl \bar{Y}_{6(n-1)-1} = -\alpha Y_{6(n-1)-2},\\
\fl \bar{Y}_{6(n-1)+1} = \alpha Y_{6(n-1)},\\
\fl \bar{Y}_{6(n-1)+3} = 0,\\
\fl \left[Y_1, \bar{Y}_{6(n-1) + 2}\right] = \alpha Y_{6(n-1)+2},\\
\fl \left[Y_2, \bar{Y}_{6(n-1)+2}  \right] = 2 \alpha^2 Y_{6(n-1)+1} + \alpha \bar{Y}_{6(n-1)+2},\\
\fl \left[Y_1, \bar{Y}_{6(n-1)+4}\right] = - \alpha Y_{6(n-1)+4},\\
\fl \left[Y_2, \bar{Y}_{6(n-1)+4} \right] = 2 \alpha^2 Y_{6(n-1)+3} - \alpha \bar{Y}_{6(n-1)+4}.  \label{Ind_014}
\end{eqnarray}
Let us prove formulas (\ref{Ind_01})--(\ref{Ind_014}) for $i = n$. We introduce the commutators of length $6n-2$:
\begin{eqnarray*}
\fl Y_{6n-1} = Y_{6(n-1)+5} = \left[Y_1, Y_{6(n-1)+4} \right],\qquad
\bar{Y}_{6n-1} = \bar{Y}_{6(n-1)+5} = \left[ Y_2, Y_{6(n-1)+4}  \right].
\end{eqnarray*}
Using the Jacobi identity and formulas (\ref{Ind_01})--(\ref{Ind_014}) we find that the following equalities hold:
\begin{eqnarray*}
\fl \left[D_x, \bar{Y}_{6n-1} \right] = \alpha^2 \left( A e^{\alpha u_0} - 2 B e^{-2 \alpha u_0}  \right) Y_{6(n-1)+3} 
- \alpha \left( A e^{\alpha u_0} + B e^{-2 \alpha u_0}\right) \bar{Y}_{6(n-1)+4} = \\
= \left[D_x, -\alpha Y_{6(n-1)+4}  \right],
\end{eqnarray*}
\begin{eqnarray*}
\fl \left[D_x, Y_{6n-1} \right] = \left[D_x, \left[Y_1, Y_{6(n-1)+4}  \right] \right]= 
 \left[Y_1, \left[D_x, Y_{6(n-1)+4}\right] \right] - \left[ Y_{6(n-1)+4}, \left[D_x, Y_1\right] \right] = \\
\fl =\left[ Y_1, \alpha \left(-A e^{\alpha u_0} + 2 B e^{-2 \alpha u_0} \right) Y_{6(n-1)+3} + \left( A e^{\alpha u_0} + B e^{-2 \alpha u_0} \right) \left[Y_2, Y_{6(n-1)+3} \right] \right] -\\
 \fl -\left[ Y_{6(n-1)+4}, \left( A e^{\alpha u_0 }  + B e^{-2 \alpha u_0}\right) Y_2 \right]= \alpha \left( -A e^{\alpha u_0} + 2 B e^{-2 \alpha u_0} \right) Y_{6(n-1)+4} + \\
\fl + \left( A e^{\alpha u_0} + B e^{-2 \alpha u_0}  \right) \left[ Y_1, \left[Y_2, Y_{6(n-1)+3} \right] \right] + \left(A e^{\alpha u_0} + B e^{-2 \alpha u_0}  \right) \left[ Y_2, Y_{6(n-1)+4} \right] = \\
\fl= \alpha \left( -A e^{\alpha u_0} + 2 B e^{-2 \alpha u_0} \right) Y_{6(n-1)+4} + \left( A e^{\alpha u_0} + B e^{-2 \alpha u_0} \right) \left[Y_2, \bar{Y}_{6(n-1)+4}  \right] + \\
\fl +  \left(A e^{\alpha u_0} + B e^{-2 \alpha u_0}  \right) \left[Y_2, Y_{6(n-1)+4}  \right] = -3 \alpha A e^{\alpha u_0} Y_{6(n-1)+4} = -3 \alpha A e^{\alpha u_0} Y_{6n-2}.
\end{eqnarray*}
Using Lemma 1 we conclude that $\bar{Y}_{6n-1} = -\alpha Y_{6(n-1)+4} $. One can see that the operator $Y_{6n-1}$ is not linearly expressed through the operators of less indices. Thus we obtain $L_{6n-1} = L_{6n-2} \oplus \left\{ Y_{6n-1}\right\}$,  $\delta(6n-1) = 1$.

Now we consider the commutators of length $6n-1$:
\begin{equation*}
Y_{6n} = \left[Y_1, Y_{6n-1}\right], \quad \bar{Y}_{6n} = \left[Y_2, Y_{6n-1} \right].
\end{equation*}
The formulas of commutation with the operator $D_x$ are:
\begin{eqnarray*}
\fl \left[ D_x, \bar{Y}_{6n} \right] = \left[ D_x, \left[Y_2, Y_{6n-1}  \right] \right] = \\
=\left[Y_2, -3\alpha A e^{\alpha u_0} Y_{6n-2}  \right] 
=-3 \alpha^2 A e^{\alpha u_0} Y_{6n-2} - 3 \alpha A e^{\alpha u_0} \left[Y_2, Y_{6n-2} \right] = \\
= -3 A \alpha^2 e^{\alpha u_0} Y_{6n-2} + 3 \alpha^2 A e^{\alpha u_0} Y_{6(n-1)+4} = 0,
\end{eqnarray*}
\begin{eqnarray*}
\fl \left[ D_x, Y_{6n} \right] = \left[D_x, \left[ Y_1, Y_{6n-1} \right]  \right] = \left[Y_1, -3 \alpha A e^{\alpha u_0} Y_{6n-2} \right]- \\
-\left[Y_{6n-1}, \left(A e^{\alpha u_0} + B e^{-2 \alpha u_0} \right)Y_2  \right] = -3 \alpha A e^{\alpha u_0} Y_{6n-1} + \\
+\left(A e^{\alpha u_0} + B e^{-2 \alpha u_0}  \right) \bar{Y}_{6n} = -3 \alpha A e^{\alpha u_0} Y_{6n-1}.
\end{eqnarray*}
According to Lemma 1, we conclude that $\bar{Y}_{6n} = 0$. The operator $Y_{6n}$ is not linearly expressed through the operators of lower order. The equalities $L_{6n} = L_{6n-1} \oplus \left\{ Y_{6n}  \right\}$, $\delta(6n) = 1$ are true.

Let us consider the commutators of length $6n$:
\begin{equation*}
Y_{6n+1} = \left[ Y_1, Y_{6n} \right], \quad \bar{Y}_{6n+1} = \left[Y_2, Y_{6n} \right].
\end{equation*}
The following formulas hold:
\begin{eqnarray*}
\fl \left[D_x, \bar{Y}_{6n+1}  \right] = \left[D_x, \left[Y_2, Y_{6n} \right]  \right] = \left[ Y_2, -3 \alpha A e^{\alpha u_0} Y_{6n-1} \right] = \\
= -3 \alpha^2 A e^{\alpha u_0} Y_{6n-1}- 3 \alpha A e^{\alpha u_0} \left[ Y_2, Y_{6n-1}  \right] = - 3 \alpha^2 A e^{\alpha u_0} Y_{6n-1}= \left[ D_x, \alpha Y_{6n} \right],
\end{eqnarray*}
\begin{eqnarray*}
\fl \left[D_x, Y_{6n+1}\right] = \left[D_x, \left[Y_1, Y_{6n} \right]\right] = \left[ Y_1, -3 \alpha A e^{\alpha u_0} Y_{6n-1}\right] 
-\left[ Y_{6n}, \left( Ae^{\alpha u_0} + B e^{-2 \alpha u_0} \right) Y_2 \right] =\\
= -3 \alpha A e^{\alpha u_0} Y_{6n} + \left( A e^{\alpha u_0} + B e^{-2 \alpha u_0} \right) \bar{Y}_{6n+1} 
= \alpha \left( -2 A e^{\alpha u_0} + B e^{-2 \alpha u_0} \right) Y_{6n}.
\end{eqnarray*}
So we have $\left[D_x, \bar{Y}_{6n+1}  \right] = \left[ D_x, \alpha Y_{6n} \right]$. Due to Lemma 1 the last equality implies that $\bar{Y}_{6n+1} = \alpha Y_{6n}$. The operator $Y_{6n+1}$ is not linearly expressed through the operators of lower order, $L_{6n+1} = L_{6n} \oplus \left\{ Y_{6n+1} \right\}$, $\delta(6n+1) = 1$.

Let us consider the commutators of length $6n+1$:
\begin{equation*}
Y_{6n+2} = \left[Y_1, Y_{6n+1}  \right], \quad \bar{Y}_{6n+2} = \left[Y_2, Y_{6n+1} \right].
\end{equation*}
For these operators the following formulas are satisfied:
\begin{eqnarray*}
\fl \left[D_x, \bar{Y}_{6n+2}\right] = \left[ D_x, \left[Y_1, Y_{6n+1}  \right] \right] = \left[ Y_1, \alpha \left(-2 A e^{\alpha u_0} + B e^{-2 \alpha u_0}  \right)Y_{6n} \right] - \\
 -\left[Y_{6n+1}, \left(A e^{\alpha u_0} + B e^{-2 \alpha u_0}  \right)Y_2  \right] 
= \alpha \left( -2 A e^{\alpha u_0} + B e^{-2 \alpha u_0} \right) Y_{6n+1} +\\
+ \left( A e^{\alpha u_0} + B e^{-2 \alpha u_0}   \right) \bar{Y}_{6n+2}.
\end{eqnarray*}
The operators $Y_{6n+2} = \left[Y_1, Y_{6n+1}\right]$ and $\bar{Y}_{6n+2} = \left[ Y_2, Y_{6n+1} \right]$ are not linearly expressed by operators of lower order. The equalities are true:  $L_{6n+2} = L_{6n+1} \oplus \left\{ Y_{6n+2}, \bar{Y}_{6n+2} \right\}$, $\delta(6n+2) = 2$.

We introduce the commutators of length $6n+2$:
\begin{equation*}
\fl Y_{6n+3} = \left[ Y_1, Y_{6n+2} \right], \quad \bar{Y}_{6n+3} = \left[ Y_2, Y_{6n+2} \right], \quad  \left[ Y_1, \bar{Y}_{6n+2} \right] , \quad \left[ Y_2, \bar{Y}_{6n+2} \right].
\end{equation*}
The following formulas are satisfied:
\begin{eqnarray*}
\fl \left[D_x, \left[ Y_2, \bar{Y}_{6n+2} \right]  \right] = \left[Y_2, \alpha^2 \left( -4 A e^{\alpha u_0} - B e^{-2 \alpha u_0}  \right) Y_{6n} \right] = \\
= \alpha^3 \left( -4 A e^{\alpha u_0} + 2 B e^{-2 \alpha u_0} \right) Y_{6n} + \alpha^2 \left( -4 A e^{\alpha u_0} - B e^{-2 \alpha u_0} \right) \left[Y_2, Y_{6n} \right] = \\
= \alpha^3 \left( -4 A e^{\alpha u_0} + 2 B e^{-2 \alpha u_0}  \right) Y_{6n} + \alpha^3 \left( -4 A e^{\alpha u_0} - B e^{-2 \alpha u_0} \right)Y_{6n} = \\
=\alpha^3 \left( -8 A e^{\alpha u_0} + B e^{-2 \alpha u_0}  \right) Y_{6n} = \left[D_x, 2 \alpha^2 Y_{6n+1} + \alpha \bar{Y}_{6n+2}  \right].
\end{eqnarray*}
So we have that $\left[D_x, \left[ Y_2, \bar{Y}_{6n+2} \right]  \right] =  \left[D_x, 2 \alpha^2 Y_{6n+1} + \alpha \bar{Y}_{6n+2}  \right]$. Apply Lemma 1 to this equality we conclude that $\left[ Y_2, \bar{Y}_{6n+2} \right]  =2 \alpha^2 Y_{6n+1} + \alpha \bar{Y}_{6n+2}  $.

Then, we find
\begin{eqnarray*}
\fl \left[ D_x, \left[Y_1, \bar{Y}_{6n+2}\right] \right] = \left[Y_1, \alpha^2 \left(  -4 A e^{\alpha u_0} - B e^{-2 \alpha u_0}\right) Y_{6n} \right] - \left[ \bar{Y}_{6n+2}, \left( A e^{\alpha u_0} + B e^{-2 \alpha u_0}  \right) Y_2\right] = \\
= \alpha^2 \left( -4 A e^{\alpha u_0} - B e^{-2 \alpha u_0} \right) Y_{6n+1} + \left( A e^{\alpha u_0} + B e^{-2 \alpha u_0} \right) \left[ Y_2, \bar{Y}_{6n+2} \right] =\\
= \alpha^2 \left( -4 A e^{\alpha u_0} - B e^{-2 \alpha u_0}  \right) Y_{6n+1} + \left(A e^{\alpha u_0} + B e^{-2 \alpha u_0}  \right) \left( 2 \alpha^2Y_{6n+1} + \alpha \bar{Y}_{6n+2} \right) = \\
= \alpha^2 \left( -2 A e^{\alpha u_0} + B e^{-2 \alpha u_0} \right)Y_{6n+1} + \alpha \left( A e^{\alpha u_0} + B e^{-2 \alpha u_0} \right) \bar{Y}_{6n+2} = \\
= \left[ D_x, \alpha Y_{6n+2} \right].
\end{eqnarray*}
Thus we have $\left[ D_x, \left[Y_1, \bar{Y}_{6n+2}\right] - \alpha Y_{6n+2} \right] $. Due to Lemma 1 this formula gives $\left[Y_1, \bar{Y}_{6n+2}\right] = \alpha Y_{6n+2}$.
Then we find
\begin{eqnarray*}
\fl \left[ D_x, \bar{Y}_{6n+3} \right] = \left[ D_x, \left[Y_2, Y_{6n+2}  \right] \right] = \\
= \left[ Y_2, \alpha \left( -2 A e^{\alpha u_0} + B e^{-2 \alpha u_0} \right)Y_{6n+1} + \left( A e^{\alpha u_0} + B e^{-2 \alpha u_0} \right)\bar{Y}_{6n+2} \right] = \\
= \alpha^2 \left( -2 A e^{\alpha u_0} - 2 B e^{-2 \alpha u_0}\right) Y_{6n+1} + \alpha \left( -2 A e^{\alpha u_0} + B e^{-2 \alpha u_0} \right) \left[Y_2, Y_{6n+1} \right] + \\
+ \alpha \left( A e^{\alpha u_0} - 2 B e^{-2 \alpha u_0} \right) \bar{Y}_{6n+2} + \left( A e^{\alpha u_0} + B e^{-2 \alpha u_0} \right)\left[Y_2, \bar{Y}_{6n+2} \right] = \\
\alpha^2 \left( -2 A e^{\alpha u_0} - 2 B e^{-2 \alpha u_0}\right) Y_{6n+1} + \alpha \left( -2 A e^{\alpha u_0} + B e^{-2 \alpha u_0} \right) \bar{Y}_{6n+2} + \\
+\alpha \left( A e^{\alpha u_0} - 2 B e^{-2 \alpha u_0} \right) \bar{Y}_{6n+2} + \left( A e^{\alpha u_0} + B e^{-2 \alpha u_0} \right) \left( 2\alpha^2 Y_{6n+1} + \alpha \bar{Y}_{6n+2}\right) = 0.
\end{eqnarray*}
It is clear by Lemma 1 that $\bar{Y}_{6n+3} = 0$. 
Now we calculate
\begin{eqnarray*}
\fl \left[ D_x, Y_{6n+3} \right] = \left[ D_x, \left[Y_1, Y_{6n+2}  \right] \right] = \left[ Y_1, \alpha \left( -2 A e^{\alpha u_0} + B e^{-2 \alpha u_0}  \right) Y_{6n+1} + \left( A e^{\alpha u_0} + B e^{-2 \alpha u_0}\right) \bar{Y}_{6n+2} \right] - \\
- \left[ Y_{6n+2}, \left( A e^{\alpha u_0} + B e^{-2 \alpha u_0} \right) Y_2\right] = \alpha \left( -2 A e^{\alpha u_0} + B e^{-2 \alpha u_0} \right) Y_{6n+2} +\\
+ \left( A e^{\alpha u_0} + B e^{-2 \alpha u_0}\right) \left[Y_1, \bar{Y}_{6n+2}  \right] + \left(A e^{\alpha u_0} + B e^{-2 \alpha u_0} \right) \left[Y_2, Y_{6n+2}  \right] = \\
=\alpha \left( -A e^{\alpha u_0} + 2 B e^{-2 \alpha u_0} \right) Y_{6n+2}.
\end{eqnarray*}
At this step we obtain $L_{6n+3} = L_{6n+2} \oplus \left\{ Y_{6n+3} \right\}$, $\delta(6n+1) = 1$.

Let us consider the commutators of length $6n+3$:
\begin{equation*}
Y_{6n+4} = \left[ Y_1, Y_{6n+3} \right], \quad \bar{Y}_{6n+4} = \left[ Y_2, Y_{6n+3} \right].
\end{equation*}
The following formulas are satisfied:
\begin{eqnarray*}
\fl \left[D_x, \bar{Y}_{6n+4}  \right] = \left[D_x, \left[Y_2, Y_{6n+3} \right]  \right] = \left[Y_2, \alpha \left( -A e^{\alpha u_0} + 2 B e^{-2 \alpha u_0}\right)Y_{6n+2}  \right] = \\
= \alpha^2 \left( -A e^{\alpha u_0} - 4 B e^{-2 \alpha u_0}  \right) Y_{6n+2} + \alpha \left( -A e^{\alpha u_0} + 2 B e^{-2 \alpha u_0} \right) \left[Y_2, Y_{6n+2}  \right] = \\
= \alpha^2 \left( -A e^{\alpha u_0} - 4 B e^{-2 \alpha u_0} \right) Y_{6n+2}.
\end{eqnarray*}
\begin{eqnarray*}
\fl \left[ D_x, Y_{6n+4} \right] = \left[ D_x, \left[Y_1, Y_{6n+3} \right] \right]
=\left[ Y_1, \alpha \left( -A e^{\alpha u_0} + 2 B e^{-2 \alpha u_0} \right) Y_{6n+2}\right] - \\
-\left[ Y_{6n+3}, \left( A e^{\alpha u_0} + 2 B e^{-2 \alpha u_0 }\right) Y_2 \right] = \\
= \alpha  \left( -A e^{\alpha u_0} + 2 B e^{-2 \alpha u_0} \right) Y_{6n+3} + \left( A e^{\alpha u_0} + B e^{-2 \alpha u_0} \right)\bar{Y}_{6n+4}.
\end{eqnarray*}
Therefore, we have $L_{6n+4} = L_{6n+3} \oplus \left\{Y_{6n+3}  \right\}$, $\delta(6n+1) = 1$. 

We consider the operators of the length $6n+4$:
\begin{equation*}
 Y_{6(n+1)-1} = \left[ Y_1, Y_{6n+4}  \right], \quad \bar{Y}_{6(n+1)-1} = \left[Y_2, Y_{6n+4}  \right], \quad \left[Y_1, \bar{Y}_{6n+4} \right], \quad \left[Y_2, \bar{Y}_{6n+4} \right].
\end{equation*}
Let us calculate the formulas by which the operator $D_x$ commutes with these operators:
\begin{eqnarray*}
\fl \left[D_x, \left[ Y_2, \bar{Y}_{6n+4} \right]\right] = \left[Y_2, \alpha^2 \left(-A e^{\alpha u_0} - 4 B e^{-2 \alpha u_0}  \right) Y_{6n+2} \right] = \\
= \alpha^3 \left(-A e^{\alpha u_0} + 8 B e^{-2 \alpha u_0}  \right)Y_{6n+2} + \alpha^2 \left(-A e^{\alpha u_0} + 8 B e^{-2 \alpha u_0}  \right) \left[Y_2, Y_{6n+2} \right] =\\
= \alpha^3 \left( -A e^{\alpha u_0} + 8 B e^{-2 \alpha u_0 }\right) Y_{6n+2} = \left[ D_x, 2 \alpha^2 Y_{6n+3} - \alpha \bar{Y}_{6n+4}\right].
\end{eqnarray*} 
Thus we have: $\left[D_x, \left[ Y_2, \bar{Y}_{6n+4} \right]\right] = \left[ D_x, 2 \alpha^2 Y_{6n+3} - \alpha \bar{Y}_{6n+4}\right]$. Due to Lemma~1 we obtain, that $\left[ Y_2, \bar{Y}_{6n+4} \right] = 2 \alpha^2 Y_{6n+3} - \alpha \bar{Y}_{6n+4}$. 

For $\bar{Y}_{6n+4}$ the following formula is satisfied:
\begin{eqnarray*}
\left[ D_x, \left[ Y_1, \bar{Y}_{6n+4} \right] \right] = \\
=\left[Y_1, \alpha^2 \left( -A e^{\alpha u_0} - 4 B e^{-2 \alpha u_0}  \right) Y_{6n+2} \right] -
- \left[ \bar{Y}_{6n+4}, \left( A e^{\alpha u_0} + B e^{-2 \alpha u_0} \right) Y_2 \right] =\\
= \alpha^2 \left( -A e^{\alpha u_0} - 4 B e^{-2 \alpha u_0} \right)Y_{6n+3} 
+ \left( A e^{\alpha u_0} + B e^{-2 \alpha u_0}  \right) \left[ Y_2, \bar{Y}_{6n+4} \right] =\\
= \alpha^2  \left(-A e^{\alpha u_0} - 4 B e^{-2 \alpha u_0}  \right) Y_{6n+3} + \left( A e^{\alpha u_0} + B e^{-2 \alpha u_0}\right) \left( 2 \alpha^2 X_{6n+3} - \alpha \bar{Y}_{6n+4} \right) = \\
= \alpha^2 \left( A e^{\alpha u_0} - 2 B e^{-2 \alpha u_0}  \right) Y_{6n+3} - \alpha \left( A e^{\alpha u_0} + B e^{-2 \alpha u_0} \right) \bar{Y}_{6n+4} = \left[ D_x, -\alpha Y_{6n+4} \right].
\end{eqnarray*}
Due to Lemma 1 we have $\left[ Y_1, \bar{Y}_{6n+4} \right] = -\alpha Y_{6n+4}$.

For $\bar{Y}_{6(n+1)-1}$ the following formula is true:
\begin{eqnarray*}
\left[D_x, \bar{Y}_{6(n+1)-1}  \right] = \left[ D_x, \left[Y_2, Y_{6n+4} \right] \right] = \\
=\left[ Y_2, \alpha \left( -A e^{\alpha u_0} + 2 B e^{-2 \alpha u_0} \right)Y_{6n+3} + \left( A e^{\alpha u_0} + B e^{-2 \alpha u_0} \right) \bar{Y}_{6n+4}\right] = \\
=\alpha^2 \left( -A e^{\alpha u_0} - 4 B e^{-2 \alpha u_0} \right)Y_{6n+3} + \alpha \left( -A e^{\alpha u_0} + 2 B e^{-2 \alpha u_0} \right) \bar{Y}_{6n+4} + \\
+ \alpha \left( A e^{\alpha u_0} - 2 B e^{-2 \alpha u_0} \right) \bar{Y}_{6n+4} + \left( A e^{\alpha u_0} + B e^{-2 \alpha u_0}  \right) \left[ Y_2, \bar{Y}_{6n+4} \right] = \\
= \alpha^2 \left( A e^{\alpha u_0} - 2 B e^{-2 \alpha u_0} \right) Y_{6n+3} + \alpha \left( -A e^{\alpha u_0} - B e^{-2 \alpha u_0} \right) \bar{Y}_{6n+4} = \left[D_x, -\alpha Y_{6n+4}  \right].
\end{eqnarray*}
Due to Lemma 1, we conclude that $\bar{Y}_{6(n+1)-1} = -\alpha Y_{6n+4}$. 

For $Y_{6(n+1)-1}$ the following formula is true:
\begin{eqnarray*}
\fl \left[D_x, Y_{6(n+1)-1}  \right] = \left[D_x, \left[Y_1, Y_{6n+4} \right]  \right] = \\
= \left[Y_1, \alpha \left(-A e^{\alpha u_0} + 2 B e^{-2 \alpha u_0}  \right)Y_{6n+3} + \left( A e^{\alpha u_0} + B e^{-2 \alpha u_0} \right) \bar{Y}_{6n+4} \right] - \\
- \left[ Y_{6n+4}, \left( A e^{\alpha u_0} + B e^{-2 \alpha u_0} \right) Y_2 \right] = \\
= \alpha \left( -A e^{\alpha u_0} + 2 B e^{-2 \alpha u_0} \right) Y_{6n+4} + \left( A e^{\alpha u_0} + B e^{-2 \alpha u_0} \right)  \left[ Y_1, \bar{Y}_{6n+4} \right] + \\
+ \left(A e^{\alpha u_0} + B e^{-2 \alpha u_0}  \right) \left[ Y_2, Y_{6n+4} \right] = -3 \alpha A e^{\alpha u_0}Y_{6n+4}.
\end{eqnarray*}
It is implies that the operator $Y_{6(n+1)-1} = Y_{1\ldots 121}$ is not linearly expressed through the operators of lower order and $L_{6(n+1)-1} = L_{6n+4} \oplus \left\{Y_{6(n+1)-1} \right\}$. Thus we have $\delta(6(n+1)-1) = 1$ and  Lemma~8 is proved. Evidently the Lemma~8 allows to complete the proof or the first part of the Theorem~2.

\subsection{The case $\alpha = -\beta$}

Now we pass to the second part of the Theorem~2. Here we prove that if polynomial $\Lambda(\lambda)$ has two different nonzero roots $\alpha$ and $\beta = -\alpha$ then the Lie-Rinehart algebra $L$ generated by the operators $X_0$ and $W_0$ is of  an infinite dimension.

First we introduce two polynomials according to the rule
\begin{equation*} 
{\Lambda}_{\alpha}(\lambda)=\frac{\Lambda(\lambda)}{\lambda-\alpha}, \quad {\Lambda}_{\beta}(\lambda)=\frac{\Lambda(\lambda)}{\lambda+ \alpha}.
\end{equation*}
Then we construct two operators $P_{\alpha}, P_{\beta} \in L$:
\begin{equation*}
P_{\alpha}={\Lambda}_{\alpha}(ad_{X_0}W_0), \quad P_{\beta}={\Lambda}_{\beta}(ad_{X_0}W_0)
\end{equation*}
and concentrate on the Lie-Rinehart algebra $L_{1} \subset L$ being a subalgebra of $L$ generated by the operators $P_{\alpha}, P_{\beta}$. By construction these operators satisfy the following commutativity relations
\begin{equation*}	
\left[ D_x, P_{\alpha}  \right] = A(u_{1}, u_{-1}) e^{\alpha u_0} X_0, \quad \left[ D_x, P_{\beta} \right] = B(u_{1}, u_{-1}) e^{- \alpha u_0} X_0,
\end{equation*}
where $A=A(u_{1}, u_{-1})$, $B=B(u_{1}, u_{-1})$ are some quasi-polynomials in $u_{1}$, $u_{-1}$. We assume that $(u_{-1}, u_1) \in D$, here $D$ is a domain in $\mathbb{C}^2$, where both $A$, $B$ do not vanish.

Let us consider the operators:
\begin{equation*}  
Y_1 = P_{\alpha} + P_{\beta}, \quad Y_2 = \frac{\partial}{\partial u}.
\end{equation*}
For these operators the following formulas are true:
\begin{equation}  \label{sin_DX1X2}
\left[D_x, Y_1 \right] = \left( A e^{\alpha u_0} + B e^{-\alpha u_0} \right) Y_2, \quad \left[D_x, Y_2\right] = 0
\end{equation}

{\bf Lemma 9.} \textit{Assume that polynomial $\Lambda(\lambda)$ defined by (\ref{Lambda_00}) has two different nonzero roots $\alpha$ and $-\alpha$. Then the following formulas hold:}
\begin{equation*}
L_{2k+1} = L_{2k} \oplus \left\{Y_{2k+1} \right\}, \quad L_{2k} = L_{2k-1} \oplus \left\{ Y_{2k}, \bar{Y}_{2k} \right\}.
\end{equation*}

{\bf Proof.} Let us consider the operator $Y_3 = \left[Y_2, Y_1\right]$. Using the Jacobi identity and formulas (\ref{sin_DX1X2}) we prove that
\begin{eqnarray*}
\fl \left[D_x, Y_3 \right] = \left[ D_x, \left[ Y_2, Y_1 \right] \right] = \left[ Y_2, \left( A e^{\alpha u_0} + B e^{-\alpha u_0} \right) Y_2 \right] = \\
= \left(A \alpha e^{\alpha u_0} - B \alpha e^{-\alpha u_0} \right) Y_2 = \alpha \left( A e^{\alpha u_0} - B e^{-\alpha u_0} \right) Y_2.
\end{eqnarray*}
We can see that $Y_3$ is not linearly expressed through the previous operators. Thus $L_3 = \left\{ Y_1, Y_2, Y_3 \right\}$. 

Let us construct the operators of lenght 3:
\begin{equation*}
Y_4 = \left[ Y_1, Y_3 \right] = Y_{121}, \qquad \bar{Y}_4 = \left[Y_2, Y_3 \right] = Y_{221}.
\end{equation*}
Using the Jakobi identity and formulas for the operators $Y_1$, $Y_2$, $Y_3$ we find:
\begin{eqnarray*}
\fl \left[ D_x, Y_4 \right] = \left[ D_x, \left[Y_1, Y_3 \right] \right] = \left[ Y_1, \alpha \left( A e^{\alpha u_0} - B e^{-\alpha u_0} \right) \right] -
\left[Y_3,  \left( A e^{\alpha u_0} + B e^{-\alpha u_0} \right)  Y_2 \right] =\\
= -\alpha \left(A e^{\alpha u_0} - B e^{-\alpha u_0}\right) Y_3 +  \left( Ae^{\alpha u_0} + B e^{-\alpha u_0} \right) \left[ Y_2, Y_3 \right] = \\
= - \alpha \left(A e^{\alpha u_0} - B e^{-\alpha u_0}\right) Y_3 + \alpha^2 \left( Ae^{\alpha u_0} + B e^{-\alpha u_0} \right) Y_1,
\end{eqnarray*}
\begin{eqnarray*}
\fl \left[ D_x, \bar{Y}_4  \right] = \left[ D_x, \left[Y_2, Y_3 \right] \right] = \left[ Y_2, \alpha \left( A e^{\alpha u_0} - B e^{-\alpha u_0} \right)Y_2 \right] = \\
= \alpha^2 \left( A e^{\alpha u_0} + B e^{-\alpha u_0}\right)Y_2 = \left[D_x, \alpha^2 Y_1  \right].
\end{eqnarray*}
Thus we obtain the equality $\left[ D_x, \bar{Y}_4 -\alpha^2 Y_1 \right] = 0 $. Due to Lemma 1 we conclude that $\bar{Y}_4 = \alpha^2 Y_1$. The operator $Y_4$ is not expressed through the operators of lower order. So $L_4 = \left\{ Y_1, Y_2, Y_3, Y_4 \right\}$. 

Now we construct the commutators of length 4:  
\begin{equation*}
Y_5 = \left[Y_1, Y_4  \right], \qquad \bar{Y}_5 = \left[Y_2, Y_4 \right].
\end{equation*}
Now we need to calculate the formulas by which the operator $D_x$ commutes with these operators:
\begin{eqnarray*}
\fl \left[ D_x, \bar{X}_5 \right] = \left[ D_x, \left[X_2, X_4  \right]  \right] = \left[ X_2, -\alpha \left( A e^{\alpha u_0} - B e^{-\alpha u_0} \right) X_3 + \alpha^2 \left( A e^{\alpha u_0} + B e^{-\alpha u_0} \right) X_1\right] = \\
-\alpha \left( A e^{\alpha u_0} - B e^{-\alpha u_0} \right) \bar{X}_4 + \alpha^3 \left( A e^{\alpha u_0} - B e^{-\alpha u_0} \right) X_1 = \\
= -\alpha \left( A e^{\alpha u_0} - B e^{-\alpha u_0} \right) \alpha^2 X_1 + \alpha^3 \left( A e^{\alpha u_0} - B e^{-\alpha u_0} \right) X_1 = 0.
\end{eqnarray*}
Based on Lemma 1 we claim that $\bar{X}_5 = 0$. We have one more formula:
\begin{eqnarray*}
\fl \left[D_x, Y_5  \right] = \left[ D_x, \left[Y_1, Y_4 \right] \right] = 
= \left[  Y_1, -\alpha \left(A e^{\alpha u_0} - B e^{-\alpha u_0}  \right) Y_3 + \alpha^2 \left( A e^{\alpha u_0} + B e^{-\alpha u_0} \right) Y_1\right] - \\
-\left[ Y_4, \left( A e^{\alpha u_0} + B e^{-\alpha u_0}  \right) X_2  \right]= \\
=  -\alpha \left(A e^{\alpha u_0} - B e^{-\alpha u_0} \right) Y_4 + \left( A e^{\alpha u_0} + B e^{-\alpha u_0} \right) \bar{Y}_5 =\\
= -\alpha \left( A e^{\alpha u_0} - B e^{-\alpha u_0} \right) Y_4.
\end{eqnarray*}
Thus we see that $Y_5$ is not linearly expressed through the operators of lower order and $L_5 = L_4 \oplus \left\{ Y_5 \right\}$. 

Now we construct the commutators of length 5:
\begin{equation*}
Y_6 = \left[Y_1, Y_5 \right], \quad \bar{Y}_6 = \left[ Y_2, Y_5 \right].
\end{equation*}
These operators are satisfied  the formulas:
\begin{eqnarray*}
\left[ D_x, Y_6 \right] = \left[ D_x, \left[ Y_1, Y_5 \right] \right] = \\
= \left[ Y_1, -\alpha \left( A e^{\alpha u_0} - B e^{-\alpha u_0}  \right) Y_4 \right] - \left[ Y_5, \left( A e^{\alpha u_0} + B e^{-\alpha u_0} \right) Y_2 \right] = \\
= - \alpha \left( A e^{\alpha u_0} - B e^{-\alpha u_0} \right) Y_5 + \left( A e^{\alpha u_0} + B e^{-\alpha u_0} \right) \bar{Y}_6.
\end{eqnarray*}
Thus we obtain that $L_6 = L_5 \oplus \left\{Y_6, \bar{Y}_6  \right\}$. 

Then we consider the commutators of length 6: 
\begin{equation*}
Y_7 = \left[ Y_1, Y_6 \right], \quad \bar{Y}_7 = \left[ Y_2, Y_6  \right], \quad \left[Y_1, \bar{Y}_6 \right], \quad \left[Y_2, \bar{Y}_6 \right]
\end{equation*}
The following formulas are true:
\begin{eqnarray*}
\fl \left[D_x, \left[ Y_2, \bar{Y}_6 \right]  \right] = \left[ Y_2, -\alpha^2 \left( A e^{\alpha u_0} + B e^{-\alpha u_0} \right) Y_4 \right] = \\
= - \alpha^3 \left( A e^{\alpha u_0} - B e^{-\alpha u_0} \right) Y_4 - \alpha^2 \left( A e^{\alpha u_0} + B e^{-\alpha u_0}  \right) \left[Y_2, Y_4 \right] = \\
= -\alpha^3 \left( A e^{\alpha u_0} - B e^{-\alpha u_0} \right) Y_4 = \left[ D_x, \alpha^2 Y_5 \right],
\end{eqnarray*}
\begin{eqnarray*}
\fl \left[ D_x, \left[ Y_1, \bar{Y}_6 \right] \right] = \left[ Y_1, -\alpha^2 \left( A e^{\alpha u_0} + B e^{-\alpha u_0} \right) Y_4\right] - \left[ \bar{Y}_6, \left(A e^{\alpha u_0} + B e^{-\alpha u_0}\right) X_2 \right] = \\
= -\alpha^2 \left( A e^{\alpha u_0} + B e^{-\alpha u_0}  \right) Y_5 + \left( A e^{\alpha u_0} + B e^{-\alpha u_0} \right) \left[Y_2, \bar{Y}_6 \right] = \\
= -\alpha^2 \left(  A e^{\alpha u_0} + B e^{-\alpha u_0}\right)Y_5 + \left( A e^{\alpha u_0} + B e^{-\alpha u_0}  \right) \alpha^2 Y_5 = 0,
\end{eqnarray*}
\begin{eqnarray*}
\fl \left[ D_x, \bar{Y}_7 \right] = \left[ D_x, \left[ Y_2, Y_6\right] \right] =\\
= \left[ Y_2, -\alpha \left( A e^{\alpha u_0} - B e^{-\alpha u_0} \right)Y_5 + \left( A e^{\alpha u_0} + B e^{-\alpha u_0}  \right) \bar{Y}_6 \right] = \\
= -\alpha^2 \left( A e^{\alpha u_0} + B e^{-\alpha u_0 } \right) Y_5 + \left( A e^{\alpha u_0} + B e^{-\alpha u_0} \right) \left[ Y_2, \bar{Y}_6 \right] = 0,
\end{eqnarray*}
\begin{eqnarray*}
\fl \left[D_x, Y_7  \right] = \left[D_x, \left[Y_1, Y_6  \right] \right] = \\
=\left[Y_1, -\alpha \left( A e^{\alpha u_0} - B e^{-\alpha u_0} \right) Y_5 + \left( A e^{\alpha u_0} + B e^{-\alpha u_0} \right)\bar{Y}_6 \right] - \\
- \left[ Y_6, \left( A e^{\alpha u_0} + B e^{-\alpha u_0} \right) Y_2 \right] = -\alpha \left( A e^{\alpha u_0} - B e^{-\alpha u_0} \right) Y_6 + \\
+ \left(A e^{\alpha u_0} + B e^{-\alpha u_0}  \right)\left[ Y_1, \bar{Y}_6 \right] + \left( A e^{\alpha u_0} + B e^{-\alpha u_0} \right) \left[Y_2, Y_6 \right] =\\
= -\alpha \left( A e^{\alpha u_0} - B e^{-\alpha u_0} \right) Y_6.
\end{eqnarray*}
Thus we see that $\bar{Y}_6 = \alpha^2 Y_5$, $\bar{Y}_6 = 0$, $\bar{Y}_7 = 0$, the operator $Y_7$ is not linearly expressed through the operators of lower order and $L_7 = L_6 \oplus \left\{Y_7 \right\}$, $\delta(7) = 1$.

It can be proved by induction that $ \left[Y_2, Y_i  \right] = 0$,
\begin{eqnarray*}
 \left[ D_x, Y_{i+1} \right] = -\alpha \left( A e^{\alpha u_0} - B e^{-\alpha u_0} \right)Y_i , \quad i=3,4,\ldots,\\
\left[ D_x, \left[ Y_2, Y_{i+1} \right] \right] = -\alpha^2 \left(A e^{\alpha u_0} + B e^{-\alpha u_0}  \right) Y_i.
\end{eqnarray*}
Then we have
\begin{eqnarray*}
\fl \left[D_x, \left[ Y_2, \left[Y_2, Y_{i+1}  \right] \right]  \right] = \left[ Y_2, -\alpha^2 \left( A e^{\alpha u_0} + B e^{-\alpha u_0} \right) Y_1 \right] = \\
= -\alpha^3 \left( A e^{\alpha u_0} - B e^{-\alpha u_0} \right) Y_i = \alpha^2 \left[ D_x, Y_{i+1} \right].
\end{eqnarray*}
This equality implies $ \left[D_x, \left[ Y_2, \left[Y_2, Y_{i+1}  \right] \right] -\alpha^2 Y_{i+1} \right]  = 0$. Using Lemma 1 we conclude that $\left[ Y_2, \left[Y_2, Y_{i+1}  \right] \right] = \alpha^2 Y_{i+1}$, $i = 4,6,\ldots,2n$.

The following formula is true
\begin{eqnarray*}
\fl \left[ D_x, \left[ Y_1, \left[Y_2, Y_{i+1}  \right] \right] \right] = \left[ Y_1, -\alpha^2 \left( A e^{\alpha u_0} + B e^{-\alpha u_0} \right)Y_i \right] - \\
- \left[ \left[ Y_2, Y_{i+1} \right], \left( A e^{\alpha u_0} + B e^{-\alpha u_0} \right)Y_2 \right] = \\
= - \alpha^2 \left( A e^{\alpha u_0} + B e^{-\alpha u_0} \right) Y_{i+1} + \left(A e^{\alpha u_0} + B e^{-\alpha u_0}  \right) \left[ Y_2, \left[ Y_2, Y_{i+1} \right] \right] = 0.
\end{eqnarray*}
Due to Lemma 1 we obtain that $\left[Y_1, \left[Y_2, Y_{i+1}  \right]  \right] = 0$, $i = 4,6,\ldots 2n$. 

Thus we conclude that $L_{2k+1} = L_{2k} \oplus \left\{Y_{2k+1} \right\} $, $L_{2k} = L_{2k-1} \oplus \left\{ Y_{2k}, \bar{Y}_{2k} \right\}$.
This completes the proof of the Lemma~9. Now the second part of Theorem~2 immediately follows from the Lemma~9.

\section{Conclusions}

In the article we study the problem of integrable classification of a rather specific but important class of two-dimensional lattices. We used to this aim the method of Darboux integrable reductions and the concept of characteristic Lie algebras \cite{H2013}-\cite{HabKuznetsova19}. By applying these implements we derived the necessary conditions of integrability for lattices of the form (\ref{eq_main}). Efficiency of these conditions is illustrated in \cite{FHKN}.

\end{document}